\input{epsf}
\documentstyle[aps,amssymb,preprint]{revtex}
\DeclareMathSymbol{\Z}{\mathbin}{AMSb}{"5A}
\begin{document}
\tightenlines

\title{Absorbing-state phase transitions with extremal dynamics}

\author{Ronald Dickman$^\dagger$ and Guilherme J. M. Garcia$^*$ \\
 {\small Departamento de F\'\i sica, Instituto de Ci\^encias Exatas}\\
 {\small Universidade Federal de Minas Gerais, Caixa Postal 702}\\
 {\small CEP 30123-970, Belo Horizonte - Minas Gerais, Brazil} }

\date{\today}

\maketitle
\vskip 0.5truecm

\begin{abstract}

Extremal dynamics represents a path to self-organized
criticality in which the order parameter is tuned to a 
value of zero.
The order parameter is associated with a phase transition to an
absorbing state.  Given a process that exhibits a phase transition
to an absorbing state, we define an ``extremal absorbing"
process, providing the link to the associated extremal (nonabsorbing)
process.  Stationary properties of the latter correspond to
those at the absorbing-state phase transition in the former.
Studying the absorbing version of an extremal dynamics
model allows to determine certain critical exponents
that are not otherwise accessible.
In the case of the Bak-Sneppen (BS)
model, the absorbing version is closely related to the
``$f$-avalanche" introduced by Paczuski, Maslov and Bak
[Phys. Rev. E {\bf 53}, 414 (1996)], or, in spreading
simulations to the ``BS branching process" also studied by
these authors.  The corresponding nonextremal process
belongs to the directed percolation universality class.
We revisit the absorbing BS model, obtaining
refined estimates for the threshold and critical
exponents in one dimension.  We also study 
an extremal version of the usual contact process,
using mean-field theory and simulation.
The extremal condition slows the spread of activity and
modifies the critical behavior radically, defining an
``extremal directed percolation"
universality class of absorbing-state phase transitions.
Asymmetric updating is a relevant perturbation for this class, even though
it is irrelevant for the corresponding nonextremal class.

\vspace{1em}
%\noindent PACS: 05.65.+b, 02.30.Ks, 05.40.-a, 87.10.+e CONFERIR
\vspace{2em}

\noindent
$^\dagger$ Electronic address: dickman@fisica.ufmg.br\\
$^*$ Electronic address: gjmg@fisica.ufmg.br

\end{abstract}

\newpage

\section{INTRODUCTION}

Extremal dynamics has been employed extensively in modelling
far from equilibrium systems such as biological
evolution \cite{BS} and driven interfaces \cite{driven,leschhorn}.
Although processes with extremal dynamics do not have a phase transition
(there is no control parameter) they
exhibit scaling properties reminiscent of those
observed at continuous phase transitions \cite{pmb94,pmb96}.
Indeed, it was suggested some time ago that the appearance of
``self-organized" scaling properties in extremal dynamics and in
sandpile models corresponds to forcing the order parameter (associated
with an underlying phase transition) to zero from above \cite{sornette95}.
The purpose of this work is to explore the connection between
these scaling properties and those observed at a 
phase transition to an absorbing state.
The connection between extremal dynamics and directed percolation (DP),
the prime example of an absorbing-state phase transition,
was first suggested by Paczuski, Maslov and Bak \cite{pmb94}
and investigated in detail by these authors in the context of the
Bak-Sneppen (BS) model and related processes \cite{pmb96}.
(The latter work, as well as Ref. \cite{Grassberger:1995}, 
clearly demonstrated that the critical
exponents of the BS model are {\it not} those of DP.)
Sornette and Dornic \cite{sornette96} and
Grassberger and Zhang \cite{gz96} have shown how a variant
of directed percolation
may be transformed via extremal dynamics to display SOC.
These studies indicate that self-organized criticality 
(SOC) \cite{BTW} under extremal
dynamics arises because the system is driven to a critical point 
associated with a phase transition to an absorbing 
state \cite{sornette96}, as is also the case for 
sandpiles \cite{Dickman et al:2000}.

In the present work we are particularly interested in the modifications
needed to transform a (non-extremal, non-SOC) model having an absorbing state
to one exhibiting SOC under extremal dynamics.  We develop a general
scheme relating the two classes of models via an intermediate,
``extremal-absorbing" process
whose absorbing-state critical point corresponds exactly to the
critical behavior observed in the corresponding SOC model.
Two examples (the
BS model and an extremal contact process) are studied in detail,
yielding refined estimates for critical properties, and evidence
of a new universality class associated with absorbing phase transitions
under extremal dynamics.

The prime example of extremal dynamics is the 
BS model \cite{BS,flyvbjerg}, proposed to explain 
mass extinctions observed in the fossil record. 
While its application in the evolutionary context is debated \cite{drossel02},
it remains an intriguing and incompletely understood example of
scaling behavior far from equilibrium.
The contact process (CP) \cite{harris} is the most familiar example 
of a Markov process exhibiting a phase transition to an 
absorbing state.  We focus on the absorbing version of the 
BS model, and the extremal version of the CP, to illustrate the relations 
between extremal dynamics and absorbing phase transitions.
Our analysis  
of the spread of activity leads 
to new or refined values for the exponents $\delta$,
$\eta$, $\nu_{||}$, $\beta'$ and $z_{sp}$, and for the
critical threshold of the BS model.  
(For extremal dynamics, the avalanche exponent 
is $\tau = 1 +\delta$.)
Finite-size scaling analysis
of stationary properties at the critical point yields
estimates of the exponent ratios $\beta/\nu_\perp$ and 
$\nu_{||}/\nu_\perp$.

Studies of modified BS models have shown that scaling properties
are insensitive to changes that preserve its basic symmetries
(that is, invariance under translation and reflection) 
\cite{Garcia and Dickman:1,Garcia and Dickman:2003,Head and Rodgers:1998},
pointing to the
existence of a BS universality class.  Nonextremal models
that exhibit a phase transition to an absorbing state,
and that possess these same symmetries, and no additional ones,  
 belong generically to
the directed percolation (DP) universality class \cite{janssen,grassberger}.
Here we show that such models fall in a new ``extremal-DP"
universality class
when modified to follow extremal dynamics.  
The absorbing phase transition corresponding to the scaling
behavior of the BS and other extremal models belongs to
the extremal-DP class, not that of ordinary
directed percolation.  In other words, extremal dynamics is a
relevant perturbation for absorbing-state transition, 
just as was shown by Sneppen in the context of
interface depinning \cite{driven}.

The balance of this paper is organized as follows.  
Section II presents a general scheme linking ordinary 
absorbing-state phase transitions and extremal dynamics
via an intermediary ``extremal-absorbing" model.
In Sec. III we describe our simulation method
and results.
Our findings regarding scaling and universality 
 are discussed in  Sec. IV.
Mean-field analyses are presented
in the Appendix.

\section{Absorbing state models and extremal dynamics}

In this section
we examine how a stochastic model with an 
absorbing-state phase transition may be transformed to 
exhibit scale invariance under extremal dynamics.
We begin, for generality, by defining a rather abstract scheme, and
then discuss specific examples.
A large class of models exhibiting an absorbing-state phase transition may
be formulated as follows \cite{liggett,hinrichsen,konno,marro}.
Consider a stochastic process ${\cal S}$ defined on a connected
graph ${\cal G}$ of $N$ sites.  (${\cal G}$ consists of a set of sites
with links between certain pairs of sites.  Typical examples are a ring of
$N$ sites, and the $d$-dimensional hypercubic lattice $\Z^d$,
with links between nearest neighbors.)
The {\it state} $\sigma(i)$ of site $i$ is 0 or 1, the latter value denoting an
active site, the former an inactive one.
For each site $i$ in ${\cal G}$ we define a neighborhood
$v(i) \subset {\cal G}$, or,
more generally, a set of neighborhoods $v_1(i)$, $v_2(i)$,...,$v_n(i)$.

The dynamics of ${\cal S}$ proceeds in steps.
Each step involves choosing an active site $i$ (the {\it central site}
for this step),
at random, and changing the states of the sites in $v(i)$ according to
a certain rule $f$.  
In case there are two or more neighborhoods, 
one of them, $v_r(i)$ say,
is chosen at random from the collection, with probability $p_r$,
and a rule $f_r$ is applied to the site or sites in $v_r$.  
In general $f$ (or $f_r$) is a probabilistic rule.  
At each step the number of active sites may change, and if at any moment there
are no active sites ($\sigma(i) = 0, \forall i \in {\cal G}$),
the process has fallen into an absorbing state and there is no further
evolution.  Otherwise the dynamics proceeds to the next step.

Using $\sigma_n$ to denote the entire set of activity variables
$\sigma(1), \sigma(2),..., \sigma(N)$ at step $n$,
the dynamics generates a sequence $\sigma_1, \sigma_2,...$ 
starting from the  initial configuration $\sigma_0$.  It is frequently of interest to
associate a continuous time variable $t$ with the process.  This is usually done by
associating a time increment $\Delta t = 1/N_a$ with each step, where $N_a$ is the
number of active sites just before the step is realized.  We define the
{\it order parameter} as $\rho(t) = \mbox{Prob} [\sigma_t(i) = 1]$, i.e., the fraction of
active sites at time $t$.  (The event space here is the set of all realizations
of the process up to time $t$, starting from a given initial 
probability distribution on configuration space.)
If the stationary order parameter, defined by 
$\lim_{t \to \infty} \lim_{N \to \infty} \rho$, vanishes, the process is said to be in
the {\it absorbing phase}; otherwise it is
in the {\it active} phase.

A simple model exhibiting a phase transition to an 
absorbing state
is the {\it contact process} (CP) \cite{harris}.  
Here we consider the one-dimensional
version.  There are three sets $v_r(i)$, conveniently denoted as
$v_0(i) = i$
and $v_{\pm}(i) = i \pm 1$.  The associated probabilitites are $p_0$
and $p_\pm = (1-p_0)/2$.  In terms of the usual 
parametrisation \cite{harris,marro}, 
$p_0 = 1/(1+\lambda)$, where $\lambda \geq 0$ represents
the rate of spread of activity.
(In the ``epidemic" interpretation of the CP, active sites represent infected
organisms, inactive sites susceptibles, and $\lambda$ is the
infection rate.)
The updating rules are: $f_0 = 0$,
$f_\pm = 1$.  In other words, an active site has a probability 
per unit time of $1/(1+\lambda)$ to become inactive, 
while an inactive site $j$ becomes active
at rate $\lambda n_a(j)/[2(1+\lambda)]$, where $n_a(j)$ is the 
number of active neighbors of site $j$.  
As is well known, the one-dimensional 
CP exhibits a continuous phase transition 
between an absorbing phase and an active one at 
$\lambda_c \simeq 3.29785$ \cite{liggett,konno,marro}.

It is convenient to associate the control 
parameter with the updating rule $f$ rather than with the 
probabilities $p_r$.  We therefore
reformulate the CP as follows.  
With the sets $v_0(i)$ and $v_\pm (i)$ defined
above, we take $p_0 = 1/2$ and $p_\pm = 1/4$,  
and define $q = \lambda/(1+\lambda)$.
The updating functions are:
\begin{equation}
f_0 = \left\{
\begin{array}{lrl}
1, & \mbox{w.p.} & q \\

0, & \mbox{w.p.} & 1-q  

\end{array} \right.
\label{f0cp}
\end{equation}
(`w.p.' denotes ``with probability"),
and
\begin{equation}
f_\pm = \left\{
\begin{array}{llll}
1, & \!\!\!\!\!\!\!\!\!\!\!\!\!\!\!\!\!\!\!\!
\!\!\!\!\!\!\!\!\!\!\!\!\!\!\!\!\!\!\!\!
\!\!\!\!\!\!\!\!\!\!\!\!\!\!\!\!\!\!\!\!
\mbox{ if } & \sigma(i\pm 1) = 1\\

\left.
\begin{array}{lrl}

1, & \mbox{w.p.} & q \\

0, & \mbox{w.p.} & 1-q 
\end{array} \right\} \mbox{ if } \sigma(i\pm 1) = 0 

\end{array} \right.
\label{fpmcp}
\end{equation}
It is easy to verify that the transition rates satisfy
$w(0 \to 1)/w( 1 \to 0) = n_a \lambda/2$, just as in the original
formulation.  The critical value $q_c \simeq 0.76733$.

The following three-site contact process (CP3) will play an
important role in our analysis \cite{jovanovic}.  For each site we define the set 
$v(i) = \{i\!-\!1,i,i\!+\!1\}$ (the central site and its nearest
neighbors).
The updating function $f$ takes values of 1 and 0 with probabilities
$q$ and $1\!-\!q$ respectively, independently at each of the three sites
in $v(i)$. (In Ref. \cite{jovanovic} this is called 'model 3'.)
Simulations of the CP3 show that it
exhibits a continuous phase transition at $q = q_c \simeq 0.63523(3)$.
Further interacting particle systems, such as the pair contact 
process \cite{iwanpcp} and the diffusive CP \cite{dcp}, can be
accomodated within the scheme set out above.

We shall assume that the process ${\cal S}$ is defined so
that the control parameters (such as $q$) are associated
with the updating rule $f$.
Each time a site is updated,
the value of $f$ may be determined
by comparing a random number $x$ with the parameter
in question.  (This is of course the usual procedure in simulations.)
In the CP3, for example, we take $f = 1$ if $x < q$, 
and zero otherwise, where $x$ is uniformly 
distributed on the interval [0,1].
Call the random number associated with the most recent updating of
site $i$, $x_i$, so that $\sigma(i) = \Theta(q - x_i)$ with $\Theta$
the unit step function.
(The initial values of the $x_i$ are assigned according to the state 
variables $\sigma(i)$.
For example, if all sites are initially active, we draw the initial
$x_i$ from the distribution uniform on $[0,q]$.)
For the CP, Eq. (\ref{f0cp}) requires that 
we update the central site $i$ with a number chosen
uniformly from [0,1].  According to Eq. (\ref{fpmcp}),
the same applies when updating an {\it inactive}
neighbor ($i \pm 1$), but when updating an {\it active} neighbor,
the random number
$x$ is drawn from the interval $[0,q]$, since an active neighbor 
remains active.

Summarizing, we have shown how a particle system ${\cal S}$ 
may be formulated using
a set of random variables $x_i$, 
such that site $i$ is active if $x_i$ is smaller than a certain
parameter $q$.  ${\cal S}$ suffers an absorbing-state phase transition
at $q = q_c$.
We now define two related processes, ${\cal S}_{EA}$ and ${\cal S}_E$.
The former, {\it extremal absorbing} process,
is obtained by modifying how the central site is
selected.  Instead of choosing it at random from among the
currently active sites, it is taken to be {\it the active site having
the smallest} $x_i$.  As in the original process ${\cal S}$,
if there are no active sites
(i.e., $x_j > q$, $\forall j \in {\cal G}$), the process has
reached an absorbing configuration and the evolution ceases.
Thus ${\cal S}_{EA}$ possesses an absorbing state, and
since the original process ${\cal S}$ exhibits a phase transition
between an active and an absorbing phase as the control parameter $q$
is varied, we expect ${\cal S}_{EA}$ to as well, 
at some value $q_{c,E}$.
(The reason is that the relative likelihood 
of generating and destroying active sites varies with $q$, just as in
${\cal S}$.)
Mean-field theory (see Appendix) yields 
$q_{c,E} = q_c$.  Due to the different
correlations generated under extremal dynamics, however, the critical 
value $q_{c,E}$ of ${\cal S}_{EA}$ is in general different from $q_c$.

We define the {\it extremal} process ${\cal S}_E$ by relaxing the condition in
${\cal S}_{EA}$, that the extremal site $i$ must be active 
(i.e., have $x_i < q$) for the dynamics to proceed.
%Since in ${\cal S}_E$ there is no distinction between active and inactive
%sites, we may dispense with the variables $\sigma(i)$.
If ${\cal S}$ is the original contact process,
then ${\cal S}_E$ is a process in which either the
minimal site or one of its nearest neighbors is updated at each step.
If ${\cal S}$ is the CP3, ${\cal S}_E$ is the familiar Bak-Sneppen
model.  Note that ${\cal S}_E$ has no absorbing state, 
hence no phase transition to such a state.  
Its stationary properties are nevertheless
intimately connected with the critical-point properties of 
${\cal S}_{EA}$, as we now explain.  

Of particular interest is the 
stationary probability density $\overline{p}(x)$ 
of site variables under extremal dynamics.  As is well known,  
$\overline{p}(x) = C \Theta (x-q_{c,E}) \Theta (1-x)$ 
in the Bak-Sneppen model, in the infinite-size limit.  
($C = 1/(1-q_{c,E})$ is the normalization factor.)
We expect $\overline{p}(x)$ to exhibit a step-function 
singularity in any extremal model ${\cal S}_E$ \cite{Garcia and Dickman:1}.
This feature is in fact already present in the original
absorbing-state model ${\cal S}$ at its critical point,
because at the critical point $q = q_c$, the stationary
density of active sites (having $x_j < q$) tends to zero as
the system size $N$ goes to infinity.  The distribution on the
``allowed" region $x > q$ is uniform, since the $x_j$ are
drawn from a uniform distribution.  
Thus $\overline{p}(x)$ jumps from zero to a finite value
at $x = q_c$.
In the {\it supercritical} regime ($q > q_c$),
$\overline{p}(x)$ is equal to a constant $p_1$ for $x < q$,
such that $qp_1 = \rho$ (the order parameter),
and takes a different constant value, $p_2$, on the interval $[q,1]$.  
Once again, the stationary density
is discontinuous at $x = q$.

What holds for ${\cal S}$ also holds qualitatively for
${\cal S}_{EA}$.  The critical value $q_{c,E}$ may, as noted, differ 
from $q_c$, but since ${\cal S}_{EA}$ exhibits an absorbing-state
phase transition, its stationary distribution $\overline{p}(x)$
also has a step-function singularity.
Just at $q = q_{c,E}$, the order parameter $\rho = 0$, but
as $N \to \infty $ the survival time of the process
tends to infinity.  This means that 
the process can survive indefinitely, with the choice of the central site 
{\it restricted to the set having} $x \leq q_{c,E}$.  
The presence 
of active sites in the range $q_{c,E} < x \leq q$ is then irrelevant, since
a site with $x \leq q_{c,E}$ is always available.  Thus for
$q \geq q_{c,E}$, the distribution $\overline{p}(x)$ exhibits a
step-function singularity at $q_{c,E}$.  Extremal dynamics
effectively ``pins" the singularity at $q_{c,E}$.
The foregoing remarks on ${\cal S}_{EA}$ obtain in the
infinite-size limit; for finite $N$ there is a nonzero probability
(for $q < 1$), that all sites have $x > q$ so that the system
eventually becomes trapped in the absorbing state.
(The mean lifetime, however, is expected to grow exponentially
with $N$, for $q > q_{c,E}$.)
	   
In summary, if ${\cal S}$ exhibits an absorbing-state phase transition,
then ${\cal S}_{EA}$ should as well, although not necessarily at the
same value of $q$.  The distribution $\overline{p}(x)$ possesses a step
function singularity in both cases.
Near the critical point 
($q \gtrsim q_c$) of the original process
${\cal S}$ we expect 
$\rho \sim (q - q_c)^\beta$, with $\beta$ the critical exponent
associated with the order parameter.  
Below the upper critical dimension 
($d_c = 4$ for DP \cite{janssen,grassberger}), $\beta < 1$.
In the supercritical regime of
${\cal S}_{EA}$, on the other hand,
$\rho = \int_0^q \overline{p}(x) dx \propto (q-q_{c,E})$ since
$\overline{p}(x)$ jumps from zero to a finite value at $x = q_{c,E}$.
Thus the order parameter exponent $\beta$ is {\it unity} in ${\cal S}_{EA}$.
This is but one example showing that the extremal condition changes 
the value of a critical exponent.  Another example, established some
time ago by Paczuski and coworkers \cite{pmb96,pmb95}, 
is that the spreading exponent
$\eta$ (defined Sec. III.C) is generically zero for extremal
dynamics, whereas $\eta > 0$ for directed percolation.
Further evidence that extremal dynamics modifies
critical exponent values will be given below.

In ${\cal S}_{E}$, the central site is always chosen (in
the $N \to \infty$ limit) from the
set $\{ i:x_i \leq q_{c,E}\}$, just as in ${\cal S}_{EA}$ at its
critical point.
Since the order parameter is zero 
in the latter case, we may assert that extremal dynamics
effectively tunes the order parameter to zero.
($\rho$ approaches zero from above as $N \to \infty$.)
As we have seen, the existence of sites with $q_{c,E} < x < q$ 
becomes irrelevant in ${\cal S}_{EA}$, in the infinite-size limit.
In other words, ${\cal S}_{E}$ is identical to the critical process
${\cal S}_{EA}$ in this limit.  (Starting from the same initial 
configuration, and using the same set of random numbers, 
the same sequence of
sites will be updated in the two processes.)

Since the extremal version of the CP3 is the familiar Bak-Sneppen model,
we shall refer to the CP3$_{EA}$ as 
the {\it absorbing Bak-Sneppen} (ABS) model.
Our objective is to characterize
the behavior of the ABS model and of the extremal and EA versions
of the contact process.  
The ABS model is closely related to the $f$-{\it avalanche} process
studied in Ref. \cite{pmb96}.  An $f$-avalanche (in the present notation,
$q$-avalanche), begins when the minimal site variable $x_{min} < q$,
the minimum having been {\it greater} than $q$ at the preceding step or steps,
and continues until the minimum is once again $> q$.  (As $q$ approaches
$q_{c,E}$ from below, the mean avalanche duration diverges.)  The dynamics of the
BS model continues, regardless of whether a given avalanche has terminated or not.
But in the ABS model $x_{min} > q$ represents an absorbing state and the
dynamics ceases.  In the BS model, it is common to analyze   
the properties of $q$-avalanches in the stationary state.  It is similarly of
interest to study stationary properties of the ABS model, attained once the
system has relaxed, after an initial transient period.  We may also study the
mean lifetime of the active state as a function of system size.  Another
approach to studying absorbing-state phase transitions consists in
following the spread of activity starting from a single active site.
This spreading phenomenon in the BS model was studied in Refs. \cite{pmb94,pmb96},
where it is called the BS branching process, and in Ref. \cite{Grassberger:1995}
under the name of the BS($\tilde{p}$) model.

The assertions regarding extremal and extremal-absorbing models
are supported by mean-field theory (MFT), which is presented in the Appendix.
In particular, for the ABS model 
$\overline{p} = \frac{3}{2} \Theta (x- \frac{1}{3})$,
just as in the MFT of the original BS model.  
In the extremal and the extremal-absorbing contact process, the
stationary probability densities exhibit (for $q > q_{c,E} = 1/2$),
{\it two} discontinuities, one at $x = \kappa \equiv q^2/(3q-1)$, the other
at $x=q > \kappa$.  [These coalesce at $q=1/2$.  Note that the
parameter $q$ continues to influence the form of $\overline{p}$
in the extremal contact process, due to the nature
of the updating rule, Eqs. (\ref{f0cp}) and (\ref{fpmcp}).]
These predictions are in qualitative agreement with
simulation.

\section{SIMULATION RESULTS}

\subsection{Absorbing Bak-Sneppen Model: Simulation Method}

Before discussing our results we insert 
an observation on simulations of
the BS model. Since the site with the smallest variable, $x_{min}$,
must be identified at each step, it becomes important to
devise an effective search strategy. An efficient general-purpose
search algorithm uses a binary tree structure to identify $x_{min}$.
One approach \cite{Grassberger:1995} utilizes a lattice of $2^n$
sites. At the first level of selection, each site is compared
with one of its neighbors and the minimum of the pair selected.
At the next level the minimum between each neighboring pair
is selected, and so on, so that at the $n$-th level the global
minimum is identified. 

A second binary scheme \cite{pmpreprint} is formulated as
follows. Site 0 is placed at the apex of the tree. Site 1
is placed on the level below the apex, to the left of 0 
if $x_1<x_0$, to the right 
if $x_1>x_0$. A site $i$ is added to the tree in the
following way: we go down the tree comparing $x_i$ with the variables
$x_1,...,x_{i-1}$, turning left or right depending on whether $x_i$ is
smaller or larger than $x_j$, until we find an empty site. 
Building the tree in this way, 
$x_{min}$ will occupy the leftmost position in the tree.
In these schemes, maintaining the tree structure, once constructed
from the initial set of variables $x_i$, requires a small number of
operations at each step, and is many times more efficient that
a repeated global search for the minimim.  We find, nonetheless,
that a suitably {\it restricted} search requires less cpu time
in the stationary state.

A special property of the BS model (shared by its absorbing version,
and by other extremal models),
is that the minimal site falls, with a probability approaching unity
as the system size grows, in the interval $[0,q_{c,E}]$.  At the same time
the density of sites with values in this interval approaches zero
as $N \to \infty$.  This suggests maintaining a list of sites having
$x < q_{c,E}$ \cite{pmnote}.  Then the search for $x_{min}$ 
may be restricted to the
list, except for the rare instances in which the latter is empty.
(For the ABS we must in any case restrict the search to sites 
with $x \leq q$.)
If the system is large, 
so that the typical number of sites with $x < q$ is not small, 
it becomes advantageous to introduce a {\it second} list, 
of sites having $x < q^{**} < q_{c,E}$.  When this
relatively short list is nonempty (as is usually the case)
the search for $x_{min}$ is restricted to it.
In studies of the BS model,
we obtained the greatest efficiency using $q^{**} = 0.54$, while the
criterion for the first list was $x < 0.65$, that is, slightly
{\it below} $q_{c,E}$.  (The occasional need to perform a global
search, in the rare instances when both lists are empty, is
more than compensated by their reduced sizes when using these 
values.)
Compared with the
binary tree method, our approach results in threefold reduction in 
CPU time, in the stationary state, for a system of 1000 sites.  
(The binary tree approach may
prove more efficient for studying transients, since initially the
lists will not be short, if the $x_i$ values are chosen uniformly
on [0,1].)

\subsection{Absorbing Bak-Sneppen Model: Stationary Properties}

Using the simulation method described above, we conducted extensive
studies of the one-dimensional 
ABS model.  We initialize the system with
all sites active and allow it to relax until mean properties
fluctuate about stationary values.  The stationary properties 
are then obtained from temporal averages over the set of surviving  
realizations.
Each step corresponds to a time interval of
$\Delta t=1/N_a$, with $N_a$ the number of active sites just
before the updating is performed.  
Results for the stationary order parameter
(i.e., the density of sites with $x < q$), on lattices
confirm that in the supercritical regime ($q > q_{c,E}$),
the order parameter grows linearly with $q - q_{c,E}$, as anticipated
in Sec. II.

We study the finite-size scaling
behavior of the stationary order parameter $\overline{\rho}$ and
of the lifetime $\tau$ at the critical point.  ($\tau$ is 
obtained from an exponential
fit to the survival probability $P_s(t)$.)  
The expected
finite-size scaling behaviors at the critical point are:
$\overline{\rho} \sim L^{-\beta/\nu_\perp}$ and 
$\tau \sim L^{\nu_{||}/\nu_\perp}$.  
We performed simulations at $q = 0.66701$ and $0.66702$,
the latter being the preferred literature value for the
threshold in the one-dimensional BS model, while the former is
favored by the results discussed in the following subsection. 
We studied systems of 1000, 2000, 4000,...,32000, sites in
simulations of $2 \times 10^7$ to $3 \times 10^8$ time steps.

For $L=4000$ - 32000, the results for the order parameter
follow a power law with 
$\beta/\nu_\perp = 0.755(5)$.  The data for smaller system sizes,
however, show systematic deviations from a pure power law, leading
us to seek a correction to scaling term;  
a correction decaying 
$\propto L^{-1/2}$ leads to a good fit.  We fit the expression
\begin{equation}
\ln \overline{\rho} = -\frac{\beta}{\nu_\perp} \ln L - \frac{b}{L^{1/2}}
\label{fitrho}
\end{equation}
to the data for $L \geq 1000$, 
allowing $\beta/\nu_\perp$ and $b$ as adjustable
parameters;  the best-fit values are $\beta/\nu_\perp = 0.769(7)$
and $b=3.69(20)$.  The simulation data, and the difference from
best fit of Eq. (\ref{fitrho}) 
are plotted in Fig. 1, showing the high quality of fit.
The data for the lifetime, using system sizes of 125, 250,...8000
yield $\nu_{||}/\nu_\perp = 2.12(1)$, with no obvious
correction term (see Fig. 2).
We also determined the stationary moment ratio 
$m = \overline{\rho^2}/\overline{\rho}^2$ at the critical
point.  The estimates for $m$ decrease
slowly with $L$, and appear, when plotted versus $L^{-0.25}$, to
approach a limiting ($L \to \infty$) value of 1.030(5) (see Fig. 3).  
(For the one-dimensional CP, $m=1.1737$ at the critical point.)
Essentially
the same results are obtained regardless of whether we use
$q_{c,E} = 0.66701$ or 0.66702 in these simulations.

\subsection{Absorbing Bak-Sneppen Model: Spread of activity}

Scaling properties at an absorbing state phase transition
are also reflected in the spread of activity from an initially
localized region \cite{torre}.  In spreading simulations of the 
ABS model we start the system with a single site $x_0 < q$
and all others above this value.  (This is completely 
equivalent to the BS branching process studied in 
\cite{pmb94,pmb96,Grassberger:1995}.)
At $q=q_{c,E}$, the process 
generates a scale-invariant pattern of activity that may be
characterized by power-laws for the survival probability $P_s(t)$,
the mean number of active sites $n(t)$ and the mean-square distance
$R^2 (t)= [n(t)]^{-1} \langle \sum_j r_j^2 (t) \rangle$.
($r_j(t)$ denotes the position of the $j-th$ active
site at time $t$.  Note that $n(t)$ is taken over all realizations,
including those that have become trapped in the absorbing state
at or before time $t$.)  The scaling laws are typically written
in the form
\begin{equation}
P_s \sim t^{-\delta}, \;\;\;\;\;\;\;
n \sim t^\eta, \;\;\;\;\;\;\;
R^2 \sim t^{z_{sp}},
\label{sprexps}
\end{equation}
relations that have been verified to high precision for various
examples \cite{hinrichsen,marro}.  (We use $z_{sp}$ to denote the spreading
exponent, to avoid confusion with the {\it dynamical} exponent $z$.)
The appearance of power laws is commonly used to locate
the critical point \cite{marro}. 

The spreading exponent $\delta$ is related to the
avalanche size exponent $\tau$, 
defined (in the BS model) via $P_D (s) \sim s^{-\tau}$,
where $P_D (s) ds$ is the probability of an avalanche having a duration
between $s$ and $s+ds$.  
Thus the survival probability $P_s(t) =
\int_t^\infty P_D(s) ds$, implying
$\tau= 1+\delta$.  

We performed spreading simulations of the ABS model at 
$q$ = 0.66699, 0.66700, 0.66701, 0.66702 and 0.66703.  Each realization was followed
up to a maximum time of about $2.7 \times 10^5$; the total number
of realizations ranged from $4 \times 10^5$ to $1.6 \times 10^6$,
depending on the value of $q$.  To locate the critical point
we study the local slope $\delta(t) = d \ln P/ d\ln t$, plotted versus
$t^{-1}$.  For $q < q_c$ the local slope is expected to veer downward
at large times, and vice-versa.  
(Numerically, $\delta(t)$ is given by the slope of a least-squares linear
fit to the data in an interval $[t_0, 20t_0]$, with geometric mean $t$.)
On the basis of the local slope
data (see Fig. 4) we conclude that $q_{c,E} = 0.66701(1)$.
This is consistent with previous estimates, which place the
threshold at 0.66702(8) \cite{Grassberger:1995} and
0.66702(3) \cite{pmb96}.
(We did not find analyses of the local slopes $\eta (t)$ or
$z_{sp} (t)$, defined analogously to $\delta (t)$, 
useful in locating the critical point.)

Analyzing the data at the critical point, we are unable to
obtain good fits to $P_s$, $n$ and $R^2$ using simple power-law
expressions.  Including a subdominant power-law correction
in the relations of Eq. (\ref{sprexps}) greatly improves the
quality of fit.  In particular, the survival
probability can be fit quite accurately using
\begin{equation}
\ln P_s \simeq  -\delta \ln t + \phi_P t^{-1/4} +C
\label{Ppwc}
\end{equation}
where $C$ is a constant and the
best-fit values are $\delta = 0.084(1)$ and $\phi_P = 0.115$.
The same value for $\delta$ is found using the data for
$q=0.66702$.
(The choice of a correction term decaying as $t^{-1/4}$ is motivated
by the fact that the local slopes $\delta(t) $ and $z_{sp}(t)$ are essentially
linear when plotted versus $t^{-1/4}$, as seen in the inset of Fig. 4.)
In Fig. 5 we plot $P_s$ and the ratio of $P_s$ to the fitting function, 
Eq. (\ref{Ppwc}); the ratio is seen to be essentially
constant for $t > 50$.
The mean-square displacement may also be fit using an asymptotic
power law and correction term.  We find
\begin{equation}
\ln R^2 \simeq z_{sp} \ln t - \phi_R t^{-1/4} +C'
\label{R2pwc}
\end{equation}
with $z_{sp} = 0.921(10)$ and $\phi_{R} = 1.703$.

It has been argued that $\eta=0$ quite generally for extremal 
dynamics \cite{pmb96,pmb95}.  Our data for the one-dimensional
ABS model support this conclusion: on a double-logarithmic plot,
$n(t)$ clearly grows more slowly than a power law.  While $\eta=0$
is compatable with $n(t)$ growing without limit as $t \to \infty$
(for example, $\propto (\ln t)^\phi$,
as suggested in Ref. \cite{Grassberger:1995}), 
our results support the conclusion
that $n(t)$ saturates at a {\it finite value} $n_\infty$ at long times.
Specifically, we are unable to fit the long-time behavior using
an expression of the form $n \sim (\ln t)^\phi$.  
On the other hand, we find $d \ln n/d \ln t \propto t^{-\omega}$, with
$\omega \simeq 0.149$, suggesting
that $n(t) \simeq n_\infty \exp [ -c t^{-\omega}]$.  In fact an excellent fit
is obtained using $c=1.92$ and $n_\infty = 14.574$, as can be seen in Fig. 6.
(Saturation of $n(t)$ does not occur on the time scale of the simulation;
for the anistropic case, shown in the inset of Fig. 6, saturation
is in fact evident.)

In the absorbing phase ($q<q_{c,E}$), 
the survival probability must vanish 
as $t \to \infty$. 
Our data follow $P_s \sim t^{-\delta}e^{-t/\tau}$, where $\tau \sim |q-q_{c,E}|^{-\nu_{||}}$, 
with $\nu_{||}=2.54(2)$. On the other hand, for $q>q_{c,E}$, the survival
probability tends to a finite value as $t \to \infty$. 
We obtain $\lim_{t\to \infty} P_s \equiv P_{\infty} \sim (q-q_{c,E})^{\beta^\prime}$, 
with ${\beta^\prime}=0.20(1)$.  (In DP and allied models $\beta' = \beta$ 
\cite{torre}, but this need not hold for models in other universality
classes.)

In the CP and other nonextremal models, the spread of activity in the
supercritical regime follows a simple pattern: the size $R$ of the active region grows
linearly with time, and the number of active sites $n$ grows $\propto t^d$.  
Our observation of
subdiffusive spreading (and saturation in $n$) at the critical point lead us to
investigate supercritical spreading in the ABS model.  We find that spreading is indeed
{\it sublinear}.  For example, using $q=0.7$ in a study extending to $t \simeq 9 \times 10^6$
to avoid transient effects, we obtain $R^2 \sim t^\chi$ with
$\chi = 1.20(4)$ and $n \sim t^\lambda$ with $\lambda = 0.61(2)$.  (The exponent 
governing $R^2$ should be twice that for $n$, since active regions have a finite
activity density in the supercritical regime.)  Once again, extremal dynamics is seen to
limit the growth of activity.

\subsection{CP3 Model}

We performed spreading simulations of the CP3 model, using the approach 
described in the previous subsection.  
Each realization is followed up to a maximum time of
$6 \times 10^4$.  Using the power-law behavior of $P_s(t)$ and $n(t)$
as the criterion for criticality, we find
$q_c=0.63525(3)$ for the CP3. 
(Note that this is some 5\% smaller than the critical value
of the corresponding extremal model.)  Analyzing the 
local slopes, we obtain
$\delta=0.162(2)$, $\eta=0.312(2)$ and $z_{sp}=1.265(4)$. These values
are fully consistent with those for
directed percolation (see Table \ref{tab1}), confirming
that the CP3 model belongs to the same universality class as the 
original contact process.

A striking difference between extremal and nonextremal
models with an absorbing state is that the spread of activity
in the critical process is much slower in the former.  This is of
course reflected in the value $\eta=0$ for extremal models,
(while for example $\eta = 0.314$ for DP in one dimension),
and in the subdiffusive growth in $R^2$ in the ABS model.
In Figs. 7 and 8 we compare typical evolutions in the ABS model and 
its nonextremal analog, the CP3, at their respective critical points.  
It is evident that 
the activity spreads much more slowly
in the ABS than in the CP3.  A further notable
difference is that in the ABS a site can remain active for a 
very long time, i.e., while the site is not the 
minimum site or a neighbor of it.  Thus the rates
of both addition and loss of active sites are much smaller in the
critical extremal process than in the corresponding nonextremal one.

\subsection{Extremal CP}

In light of the discussion of Sec. II, it is of interest to
study the behavior of other absorbing-state models 
under extremal dynamics.  As a first step we report simulation results
for the extremal-absorbing contact process (CP$_{EA}$). 
We performed spreading simulations to determine $q_{c,E}$ and
the exponents $\delta$, $\eta$ and $z_{sp}$, using simulations
running to a maximum time of $6 \times 10^4$ in $5 \times 10^5$
independent realizations.  We find $q_{c,E} = 0.79415(5)$ for
the extremal CP, compared with 0.76733 for the original (non-extremal)
process.  (Note that, as in the comparison between the CP3 and ABS
models, $q_{c,E} > q_c$.  This may reflect the slower spread of activity 
under extremal dynamics.)

As in the case of the
ABS model, the decay of the survival probability at the critical
point follows an expression of the form of Eq. (\ref{Ppwc}), here with best-fit parameters
$\delta=0.0855$ and $\phi_P = 0.226$.  The exponent $\delta$ is essentially the
same as found for the ABS model, while the 
correction term is about twice as large.  At the critical point
the derivative $d \ln n/d \ln t \sim t^{-0.1}$, again indicating a behavior
of the form $n(t) \simeq n_\infty \exp [ -c t^{-\omega}]$, 
here with $\omega = 0.1$.  The mean-square distance of active sites
from the origin grows in a manner similar to that in the ABS model.
We are again able to fit the data for $R^2$ using an expression of the form of 
Eq. (\ref{R2pwc}), with
$z_{sp} = 0.932$ and $\phi_{R^2} = 2.026$.
These results strongly suggest that the CP$_{EA}$ belongs to the same 
universality class as the ABS model. 

We turn now to the rather surprising behavior of the stationary
probability density $\overline{p}(x)$ in the extremal CP.  Recall that
mean-field theory predicts $\overline{p}(x) = 2 \Theta (x-1/2)$ for 
$q < q_{c,E} = 1/2$, while for $q > 1/2$ there are two steps, one 
at $x = \kappa \equiv q^2/(3q-1)$, the other at $x=q$.
In simulations of the CP$_E$ on a ring we find a single step
discontinuity for $q < q_{c,E} = 0.79415$, and, for $q>q_{c,E}$, a pair of steps, one at
$x=q$, the other at $x=q_s<q$.  The positions of the
singularities as obtained in simulation (using data for system sizes
$L=100$, 200,...,1600 to extrapolate the position in the infinite-size
limit), are shown in Fig. 9.  The lower singularity
$q_s$ is seen to bifurcate from $q=q_{c,E}$ just at the critical point, 
in qualitative agreement with MFT.  Note however that the
position of the singularity is not constant for $q < q_{c,E}$, as 
predicted by MFT.  

The density $\overline{p}(x)$ is shown for
$q = 0.794 \simeq q_{c,E}$ and $q=0.85$ in Fig. 10.  In the latter case
it is evident that the step at $x=q$ is sharp (this is true even for small
systems) and derives from the singular nature of the updating rule.
The step at $x=q_s$, by contrast, is subject to finite-size rounding, and
becomes sharper with increasing system size, as is characteristic of
a critical singularity.  The finite width of the peak at $x=q_c$
(in the process with $q =  q_{c,E}$),
 appears to be a finite size effect as well:
it becomes sharper with increasing $L$,
suggesting that the singularities merge in the limit $L \to \infty$.

\subsection{Anisotropic ABS Model}

The scaling behavior of the Bak-Sneppen model changes when the
updating rule is asymmetric \cite{vendruscolo}.  The same critical exponents
are found for 
a highly anisotropic version in which at each step, only the
minimal site and its neighbor on the right are updated \cite{Maslov:1998},
and for weak anisotropy \cite{Garcia and Dickman:2003,head}, so that one may
identify an anisotropic BS universality class.  In this section
we report results of spreading simulations of the anisotropic
absorbing BS model.
To obtain these results, we simulated the anisotropic ABS 
(in the highly anisotropic version) in studies extending to a
maximum time of 1.6$\times 10^5$, using 3$\times 10^5$ realizations.

Analysing the local slope $\delta(t) = d \ln P/ d\ln t$, 
we determined the threshold of this model to be $q_{c,E} =0.72370(2)$.
(The best previous estimate is $0.7240(1)$
\cite{Garcia and Dickman:2}.)
A typical evolution of the critical spreading process is shown in Fig. 11.
The local slope $\delta(t)$ yields the estimate $\delta = 0.234(5)$.
For anisotropic models we define $R^2 (t)$ as the mean-square {\it radius
of gyration}, i.e., the distance is measured relative to the current
center of mass of the set of active sites, rather than to a fixed origin.
This is done to eliminate a spurious contribution due simply to the overall
drift in the active region.
For the anisotropic ABS model $R^2 (t)$ may again be fit with an expression
of the form of Eq. (\ref{R2pwc}), with
$z_{sp}=1.425(10)$ and $\phi_{R}=2.3(2)$. 
The exponents $\delta$ and $z_{sp}$ are quite different from those of the isotropic model.
Despite these differences,
we again find $\eta=0$ for the anisotropic model.
As before, the mean number of active sites $n(t)$ saturates at long times,
more rapidly in fact than in the isotropic ABS model (see Fig. 6, inset).  
We are able to
fit the data well using 
$n(t)=n_{\infty}[1- e^{-ct^{1/4}}]$ with parameters $n_\infty = 5.206(3)$
and $c = 0.348$.

The nonextremal model corresponding to the anisotropic ABS model is a two-site
contact process, CP2, which is simply the CP3 with updating restricted to the
central site and its neighbor on the right.  We have verified that the spreading exponents
of the CP2 model are those of directed percolation.  (Here again, we define $R^2$
as the mean-square radius of gyration.)  This leads to the interesting conclusion that a
perturbation (asymmetric updating) that is {\it irrelevant} for a nonextremal model
is relevant for the corresponding extremal system.
(We note that, because the two sites in the CP2 are updated in the same manner,
the model does not fall in the so-called anisotropic-DP class, for which bonds
along different axes are present with different probabilities \cite{jaff85}.)

\section{CONCLUSIONS}

We investigate the relation between extremal dynamics,
exemplified by the Bak-Sneppen model, and nonextremal models exhibiting a
phase transition to an absorbing state, using general arguments,  
mean-field theory and simulation.  The relation between
the BS model and directed percolation was already suggested some
time ago \cite{pmb94,pmb96}.  Here we clarify this
connection by showing how a generic
absorbing-state model can be transformed to an extremal one via the
associated extremal-absorbing model.  The nonextremal
`precursor' of the BS model is a three-site contact 
process \cite{jovanovic}, CP3,
which, like the original CP, belongs to the directed percolation
universality class.  The BS model and the extremal
version of the CP belong to a common universality class that may be dubbed
`extremal DP' (EDP).  A number of extremal dynamics classes distinct
from EDP are discussed in \cite{pmb96};
another example is the anisotropic BS model. 
We expect that further extremal dynamics universality
classes exist, for example an extremal parity-conserving class \cite{paritycons},
although examples of the latter have yet to be studied.  

Our results for the critical exponents of the EDP class, which 
includes the BS model and the extremal CP, are compared against those
of ordinary DP (in one spatial dimension) in Table I.  (Here we have
taken the values $\eta=0$ and $\beta=1$ to be {\it exact} for EDP.)
The differences between the two sets of exponent values are evident.
Our results $\tau = 1+ \delta = 1.084(1)$, and 
$z = \nu_{||}/\nu_\perp =   2.12(1)$ are in agreement
with the earlier estimates \cite{pmb96} of 1.07(1) and 2.10(5),
respectively.  Our result is however somewhat higher than
Grassberger's result $\tau = 1.073(3)$ \cite{Grassberger:1995}.

Certain scaling relations are expected to hold among the critical 
exponents \cite{hinrichsen,marro,torre}.
In spreading processes one expects 
$z_{sp} = 2 \nu_\perp/\nu_{||}$; our data are nearly consistent with this, yielding 
$2 \nu_\perp/\nu_{||} - z_{sp} = 0.022(14)$.
The relation $\beta' = \delta \nu_{||}$ is also
satisfied: our data yield $\beta' - \delta \nu_{||} = -0.013(14)$.
Finally, we consider the generalized hyperscaling relation \cite{mendes}
\begin{equation}
2\left(1 + \frac{\beta}{\beta'} \right) \delta + 2\eta = dz_{sp}  \;,
\label{ghypsc}
\end{equation}
in $d$ dimensions.
Using our data, we find the difference between the two sides of this
relation to be 0.09(6).  
Our results are marred by another inconsistency that may reflect
corrections to scaling or finite size effects:
the product $(\beta/\nu_\perp)^{-1} (\nu_{||}/\nu_\perp) \nu_{||}^{-1}$,
with the first two factors determined from finite-size scaling
at the critical point, and the final factor obtained from the
decay of the survival probability in the subcritical regime,
should equal $\beta=1$; our data yield 1.08(3).
These minor inconsistencies 
suggest that one or more of the exponents may be in error by 5\% or so.
Refining their values will require accumulating larger data sets
in simulations of larger systems, a task we leave for future work.
(The studies reported here were quite demanding
computationally, representing approximately 6 months' cpu time
on an alpha workstation.)

In the course of our study we revisit a three-site contact process
(CP3) that is the nonextremal analog of the BS model \cite{jovanovic}.  We verify that the
CP3 belongs to the universality class of directed percolation, 
as expected \cite{Grassberger:1995}.  Similarly,
we define extremal and extremal-absorbing versions of the original contact
process (CP$_E$ and CP$_{EA}$, respectively) and verify that their scaling properties
are the same as those of the BS model.  The stationary probability density
for the CP$_E$ follows, in general terms, the predictions of mean-field theory,
but certain interesting differences exist, as detailed in Sec. III.E.

It is clear that when an absorbing-state model is modified to follow
extremal dynamics, its critical exponents are altered.  
In general extremal dynamics tens to slow the spread of activity
in the critical and supercritical regimes. 
One may nevertheless
inquire whether any more general features of the original model
are preserved under `extremalization'.  A candidate for such a
conserved property is the critical dimension $d_c$.  In critical phenomena,
various universality classes (differing in the symmetry group of the 
order parameter, or the presence of conserved quantities) may
share the same $d_c$ if the algebraic structure of their
continuum description (in particular, the power of the lowest-order
nonlinear term in the order parameter, in a Landau-Ginzburg-Wilson
effective Hamiltonian) is the same.  Thus $d_c = 4$ for all models in the $n$-vector
family.  Extending this idea to extremal models is
questionable, since there is no continuum description at hand.
(At first glance, the notion of extremal dynamics in a
description using a {\it continuous} activity density seems
problematic, since there is always one and only one 
extremal site.)  Be that as it may, it seems plausible that
if the field theory for DP 
\cite{janssen,grassberger,cardy} were somehow modified to reflect extremal dynamics, 
the dominant nonlinearity would not change, so that $d_c$ would
retain its value of four, as in DP.  The upper critical 
dimension $d_c=4$ for the BS model was established some time ago by Boettcher
and Paczuski \cite{Boettcher:2000}.  
Our argument suggests that extremal versions of other absorbing-state
models have the same upper
critical dimension as the corresponding nonextremal model.
We hope to test this prediction in future work.

Studying the anisotropic ABS model and its nonextremal counterpart, the CP2 model, we
find that anisotropy is a relevant perturbation for extremal DP, while
it is irrelevant for the corresponding nonextremal class.  We suspect that other 
perturbations, such as diffusion, may exhibit a similar pattern of relevance.

\vspace{1cm}
\noindent {\bf ACKNOWLEDGMENTS}
\smallskip

\noindent We are grateful to Stefano Zapperi and Jafferson Kamphorst Leal da Silva
for informative
discussions, and to Paulo M. C. de Oliveira for helpful comments on the
manuscript.
We acknowledge CNPq and FAPEMIG, Brazil, for financial support.

\newpage
\centerline
{\bf APPENDIX: MEAN-FIELD THEORY}

\subsection{Extremal dynamics as a zero-temperature limit}

There are several ways of formulating a mean-field theory (MFT)
for extremal models.  First we consider
an approach \cite{Dickman et al:2000,head} in which the probability of a 
site $i$ being chosen as the central site is proportional to 
$e^{-\beta x_i}$; 
extremal dynamics is recovered in the limit $\beta \to \infty$.
(In the present discussion the parameter $\beta$ 
bears no relation to the critical exponent denoted by the same symbol in the
main text.)  
Applied to the BS model, this approach yields the distribution 
$\overline{p}(x) = (3/2) \Theta (x - 1/3) \Theta (1-x)$
when $\beta \to \infty$ \cite{Dickman et al:2000,head}.  

In the ABS model the distribution $p(x)$ evolves via
\begin{equation}
\frac{\partial p(x,t)}{\partial t} = - e^{-\beta x} p(x) \Theta(q-x)
 + 3 \int_0^q e^{-\beta y} p(y,t) dy - 2p(x,t) \int_0^q e^{-\beta y} p(y,t) dy
\label{mf1}
\end{equation}
The first term represents a site with value $x$ being selected as the central site, 
which is only possible if $x < q$. The second term
reflects updating three sites with new variables uniform
on $[0,1]$, with the integral representing the overall rate
of events.  The final term represents updating of the two
neighboring sites, and is obtained using the mean-field factorization
of the nearest-neighbor joint probability density: $p(x,y,t) \simeq p(x,t)p(y,t)$.
Note that in writing Eq. (\ref{mf1}) we have associated a time increment
$dt = 1/N$, with $N$ the number of sites, with each event.

Eq. (\ref{mf1}) admits an infinite set of stationary solutions for
which $\overline{p}(x) = 0 $ on the interval $0 \leq x \leq q$.  These represent absorbing states.
To seek an active stationary solution we let
$I= \int_0^q e^{-\beta y} p(y,t) dy$, and equate the time derivative to zero,
yielding
\begin{equation}
\overline{p}(x) = \frac{3I}{2I + \Theta(q-x) e^{-\beta x}}
\label{ss1}
\end{equation}
To determine $I$ we multiply Eq. (\ref{ss1}) 
by $e^{-\beta x}$ and integrate from $x=0$ to $x=q$,
leading to $I = (e^{-\beta/3} - e^{-\beta q})/[2(1 - e^{-\beta/3})]$, 
so that
\begin{equation}
\overline{p}(x) = \frac{\frac{3}{2} (e^{-\beta/3} - e^{-\beta q})}
{(e^{-\beta/3} - e^{-\beta q}) + \Theta(q-x) e^{-\beta x} (1 - e^{-\beta/3})}
\label{mfsta}
\end{equation}
In the limit $\beta \to \infty$,
we find, for $q > 1/3$,  
the singular density $\overline{p}(x) = (3/2) \Theta (x-1/3)$.
This is precisely the MF result for the original BS model.  
[When we take $\beta \to \infty$, the above expression reduces to
$\overline{p}(x) = (3/2) \Theta (x-q)$ for $q < 1/3$.  But this
density is not normalized on $[0,1]$ and so must be rejected.
We are left with only absorbing stationary solutions for $q < 1/3$.]
Thus $q_{c,E} = 1/3$ in the MFT of the absorbing Bak-Sneppen
model.  Note that
the parameter $q$ is irrelevant
for $q > q_{c,E} = 1/3$, as was argued in Sec. II.

A moment's reflection shows that the evolution of $p(x)$
in the (nonextremal) CP3 model is given by Eq. (\ref{mf1}) with $\beta = 0$,
since all active sites are then equally likely to be chosen as the
central site.  Taking the limit $\beta \to 0$ of the stationary
solution, Eq. (\ref{mfsta}), one finds, for $q \geq 1/3$, 
the stationary density
\begin{equation}
\overline{p}(x) = \left\{
\begin{array}{lr}
\frac{1}{2}(3 - q^{-1}), & x < q \\
\\
\frac{3}{2}, & q <x \leq 1  

\end{array} \right.
\label{cp3st}
\end{equation}
Eq. (\ref{cp3st}) confirms that the stationary density of 
a (nonextremal) model exhibiting an absorbing state phase transition
is characterized by a steplike singularity, as asserted in Sec. II.
For $q<1/3$, Eq. (\ref{cp3st}) yields an unphysical, negative 
density, showing that $q_c = 1/3$ for the CP3, in the mean-field
approximation.

The foregoing analysis is readily extended to the extremal-absorbing contact
process (CP$_{EA}$) defined in Sec.~II.  The rate of events is again
given by $I= \int_0^q e^{-\beta x} p(x) dx$.  At each event,
there is a probability of 1/2 that the central site (which must
have $x < q$) is replaced, while with probability 1/2 a
neighbor is updated.  Thus the loss terms in the equation for
$p(x,t)$ are: $-(1/2)[e^{-\beta x} \Theta(q-x) + I]p(x)$.
The gain term corresponding to updating of the central site
is simply $I/2$, but for updating a neighbor it is
$(I/2)[1-P(q) + (1/q)\Theta(q-x) P(q)]$, where
$P(x) = \int_0^x p(y) dy$ is the probability that a given
site $i$ has $x_i < x$.  (The reason is
that when updating an {\it active} neighbor, the new variable
is chosen from the distribution uniform on $[0,q]$.)  Thus
the MF equation of motion is

\begin{equation}
\frac{\partial p}{\partial t} = -\frac{1}{2} e^{-\beta x} p(x) \Theta(q-x)
 + \frac{I}{2} \left[ 2 - P(q) + \frac{\Theta(q-x)}{q}P(q) - p(x) \right]
\label{mf2}
\end{equation}
To find the stationary solution $\overline{p}(x)$ we first 
note that for $0 \leq x < q$,
setting $\partial p/\partial t $ to zero yields 
\begin{equation}
\overline{p}(x) = AI/(I+e^{-\beta x})
\label{pxs}
\end{equation}
where $A = 2 + (q^{-1}-1)P(q)$.  Integrating Eq. (\ref{pxs})
from $x=0$ to $x=q$, we find
\begin{equation}
P(q) = Aq - \frac{A}{\beta}\int_{e^{-\beta q}}^1 \frac{du}{I+u}
\label{pqss}
\end{equation}
If we now multiply Eq. (\ref{pxs}) by $e^{-\beta x}$ and integrate
over the same interval, we obtain
\begin{equation}
\frac{A}{\beta}\int_{e^{-\beta q}}^1 \frac{du}{I+u} = 1
\label{int}
\end{equation}
leading to $P(q) = 2 - q^{-1}$ in the stationary state.  We see
that $q_{c,E} = 1/2$ as in the MFT of the original contact process.
Evaluating the integral in Eq. (\ref{int}) one finds
$I = (e^{-\beta \kappa} - e^{-\beta q})/(1 - e^{-\beta \kappa})$,
where $\kappa = q^2/(3q-1)$.  The stationary density is 
\begin{equation}
\overline{p}(x) = \frac{1}{q} 
\frac{(e^{-\beta \kappa} - e^{-\beta q})[1 +(2-q^{-1})\Theta(q-x)]}
{(e^{-\beta \kappa} - e^{-\beta q}) 
+ e^{-\beta x}(1 - e^{-\beta \kappa}) \Theta(q-x) }
\label{pxscp}
\end{equation}
For $q > 1/2$, we have $\kappa < q$, and in the limit $\beta \to \infty$,
\begin{equation}
\overline{p}(x) = \left\{
\begin{array}{lr}
0, & x < \kappa \\
\\
\frac{1}{\kappa}, & \kappa < x < q  \\
\\
\frac{1}{q}, & x > q 
\end{array} \right.
\label{pxscpext}
\end{equation}
which is normalized and exhibits step-function singularities
at $x = \kappa$ and $x = q$.
For $q < 1/2$ on the other hand, $\kappa > q$ and Eq.(\ref{pxscp})
does not yield an acceptable probability density, and we conclude that
the only stationary state is the absorbing one.
The critical point of the CP$_{EA}$ thus falls at $q_{c,E} = 1/2$
in MF approximation.

Taking $\beta \to 0$ in Eq. (\ref{pxscp}), we obtain the probability density for
the original CP:
\begin{equation}
\overline{p}(x) = \left\{
\begin{array}{lr}
\frac{2q-1}{q^2}, &  x < q  \\

\frac{1}{q}, & x > q 
\end{array} \right.
\label{pxscp0}
\end{equation}

Finally, for the extremal CP, the equation of motion is
\begin{equation}
\frac{\partial p}{\partial t} = -\frac{1}{2} e^{-\beta x} p(x)
 + \frac{I}{2} \left[ 2 - P(q) + \frac{\Theta(q-x)}{q}P(q) - p(x) \right]
\label{mf3}
\end{equation}
with $I = \int_0^1 e^{-\beta x} p(x) dx$.
To find the stationary solution we write
\begin{equation}
\overline{p}(x) = 
\frac{2 + [q^{-1} \Theta(q-x) - 1]P(q)}{1 + I^{-1}e^{-\beta x}}
\label{cpxest}
\end{equation}
Integrating from $0$ to $q$ and solving for $P(q)$ we find
\begin{equation}
P(q) = \frac{2(q-\gamma)}{q+ \gamma (q^{-1}-1)}
\label{pqcpx}
\end{equation}
where
\begin{equation}
\gamma = \frac{1}{\beta} \ln \frac{I+1}{I+e^{-\beta q}}
\label{gamma}
\end{equation}
Now multiply Eq. (\ref{cpxest}) by $e^{-\beta x}$ and integrate from
0 to 1 to obtain
\begin{equation}
1 = A \int_0^q \frac{e^{-\beta x} dx}{I + e^{-\beta x}}
   + A' \int_q^1 \frac{e^{-\beta x} dx}{I + e^{-\beta x}}
\label{one}
\end{equation}
where $A = 2/[q+ \gamma (q^{-1}-1)]$ and $A' = \gamma A/q$.
If $P(q) > 0$, the first term on the r.h.s. of Eq. (\ref{one})
is nonzero and dominates as $\beta \to \infty$.  Equating the
first term to unity then leads to $\gamma = q^2/(3q-1) = \kappa$,
and thence to $P(q) = 2 - q^{-1}$ which is positive for $q > 1/2$.
A simple calculation then yields the distribution of Eq. (\ref{pxscpext})
in the limit $\beta \to \infty$.

If $q < 1/2$ the above solution is not valid since it implies $P(q) < 0$.
We therefore take $P(q) = 0$, implying $\gamma = q$, and so $A = A' = 2$.
Eq. (\ref{one}) now reads
\begin{equation}
1 = 2q 
   + \frac{2}{\beta} \ln \frac{I+e^{-\beta q}}{I+e^{-\beta}}
\label{onea}
\end{equation}
Solving for $I$ and inserting the result in Eq. (\ref{cpxest}), we
find in this case 
$\lim_{\beta \to \infty} \overline{p}(x)= 2 \Theta(x-1/2)$.  
These results have been verified via numerical integration.

\subsection{Extremal dynamics on a complete graph}

Another approach to formulating MFT for the BS model considers
extremal dynamics on an $N$-site complete graph or 
random-neighbor model (two neighbors are selected at random each time 
a site is updated); the stationary density $\overline{p}(x)$
becomes a step function in the infinite-size limit 
\cite{flyvbjerg,Garcia and Dickman:2003,deBoer:1994,deBoer:1995}.
We now extend this approach to the ABS model.
Let $P(x) = \mbox{Prob} [x_i < x] = \int_0^x p(y) dy$ be the 
distribution function and let $Q(x) = 1 - P(x)$.
By definition $Q$ is a non-increasing function with $Q(0) = 1$ and 
$Q(1) = 0$, since $p(x) = 0$ outside the interval [0,1].

Activity in the ABS model is predicated on the minimal site
$x_{min}$ being smaller than $q$; the probability of
this event, under the MF factorization, is $1 - [Q(q)]^N$.
Given $x_{min} < q$, updating the extremal site and two
neighbors results, on the average, in the increment:
$dP(x) = (1/N)\{-[1 - Q(x_<)^N] - 2P(x) + 3x \}$,
where $x_< \equiv \min\{x, q\}$, so that the first term
represents loss of the minimal site, the second removal
of two neighbors, and the third random replacement
of the three site variables with numbers uniform on [0,1].
If we adopt a time increment $dt = 1/N$ for each such
event, the equation of motion for $P$ is
\begin{equation}
\frac{\partial P(x,t)}{\partial t} = - [1 - Q(x_<,t)^N] 
+[1 - Q(q,t)^N][3x - 2P(x,t)]
\label{dPdt1}
\end{equation}
Note that
the evolution ceases if $Q(q,t) = 1$, i.e., if there are no
active sites.  (Since $Q$ is nonincreasing $Q(q) = 1
\Rightarrow Q(x_<) = 1$.)

For $q > 1/3$, the stationary solution to Eq. (\ref{dPdt1}) corresponds 
to a density
$\overline{p}(x)$ that approaches a step function,
$(3/2) \Theta (x-1/3)$, as $N \to \infty$.  
A simple calculation yields the dominant contribution for large $N$:
\begin{equation}
\overline{Q} \simeq \left\{
\begin{array}{lr}
(1-3x)^{1/N}, & x < \frac{1}{3} \\

\frac{3}{2} (1-x) + {\cal O}(e^{-\mbox{const.}N}) & x >  \frac{1}{3} 

\end{array} \right.
\label{Qss}
\end{equation}
(One should note however that the convergence is nonuniform in $x$,
being slower the closer $x$ is to the critical value of 1/3.)
For $q < 1/3$ we are unable to find an accepable stationary solution with
$Q < 1$ (i.e., $\overline{p} > 0$), for $x < q$, and conclude that only absorbing
solutions exist.

The analysis of the ABS model on a complete graph confirms that in 
the infinite-size limit,
the model enjoys the usual properties of the BS model for 
$q > q_{c,E} = 1/3$, and
falls into the absorbing state for $q < 1/3$.

The evolution of $P(x,t)$ in the extremal CP follows, in MF approximation, 
the equation
\begin{equation}
\frac{\partial P(x,t)}{\partial t} = - \frac{1}{2}[1 - Q(x,t)^N] +\frac{x}{2} 
- \frac{1}{2}P(x,t) + \frac{1}{2} [x^* P(q,t) + x Q(p,t) ] 
\label{dPdt2}
\end{equation}
where $x^* = \min\{x/q,1\}$.  Numerical integration shows that the 
solution converges, for large $N$, to a stationary distribution 
consistent with the singular density
found above in the limit $\beta \to \infty$.

\newpage

\begin{table}
\caption{Critical exponents for the one-dimensional absorbing Bak-Sneppen model (ABS) and 
contact process (CP).
CP exponents from Refs. [23] and [41].}
\begin{center}
\begin{tabular}{|c|c|c|}
Exponent            &    ABS     &     CP     \\
\hline
$\beta$             &     1      & 0.27649(4) \\
$\beta^\prime$      &   0.20(1)  & ($=\beta$) \\
$\nu_{||}$          &   2.54(2)  & 1.73383(3) \\
$\beta/\nu_{\perp}$ &   0.77(1)  & 0.25208(5) \\
$\nu_{||}/\nu_\perp$ &   2.12(1) & 1.58071(11)\\
$\delta$            & 0.084(1)   & 0.15947(3) \\
$\eta$              &     0      & 0.31368(4) \\
$z_{sp}$            & 0.921(10)  & 1.26523(3) 
\label{tab1}
\end{tabular}
\end{center}
\end{table}

\begin{table}
\caption{Spreading exponents for the CP3 and CP$_{EA}$ models and the anisotropic absorbing 
Bak-Sneppen (AABS)
model in one dimension.}
\begin{center}
\begin{tabular}{|c|c|c|c|}
Exponent       &    CP3     & CP$_{EA}$  &   AABS      \\
\hline
$\delta$       &  0.162(2)  & 0.0855(20) &  0.234(5) \\
$\eta$         &  0.312(2)  &   0        &  0        \\
$z_{sp}$       &  1.265(4)  & 0.932(20)  &  1.425(10)
\label{tab2}
\end{tabular}
\end{center}
\end{table}

%\begin{figure}
%\epsfysize=6cm
%\epsfxsize=6cm
%\centerline{
%\epsfbox{perfil.eps}}
%\caption{Stationary probability densities for the BSab with $n=5$.}
%\label{fig1}
%\end{figure}

\newpage
\noindent FIGURE CAPTIONS
\vspace{1em}

%rfss.eps
\noindent FIG. 1.  Stationary activity density (filled symbols)
$\overline{\rho}$ versus system size $L$ in the one-dimensional
ABS model at the critical point.  Open symbols: difference
between $\overline{\rho}$ and the fitting function, Eq. (\ref{fitrho}),
shifted vertically for visibility.
\vspace{1em}

%tauvl.eps
\noindent FIG. 2. Mean lifetime $\tau$ (filled symbols) versus system size $L$
in the one-dimensional
ABS model at the critical point.  
Open symbols: $\tau/L^{\nu_{||}/\nu_\perp}$ (shifted vertically
for visibility).
\vspace{1em}

%mfss.eps
\noindent FIG. 3.  Moment ratio $m$ for the ABS model versus system size
$L^{-0.25}$.  Points: simulation data; line: best linear fit, 
$m = 1.0295 + 0.268 L^{-0.25}$.
\vspace{1em}

%deltat.eps
\noindent FIG. 4. Local slope $\delta(t)$ versus $1/t$ in the ABS model.
$q$ values (bottom to top) 0.66700, 0.66701, 0.66702 and 0.66703.
Inset: data for $q=0.66701$ plotted versus $1/t^{0.25}$.
\vspace{1em}

%p701.eps
\noindent FIG. 5. Survival probability $P_s(t)$ in the ABS model at the critical
point, $q= 0.66701$.  The nearly constant function represents
the ratio of $P_s$ to the fitting function, Eq. (\ref{Ppwc}).
\vspace{1em}

%n701.eps
\noindent FIG. 6. Mean number of active sites $n(t)$ in the ABS model at the critical
point, $q= 0.66701$.  
The solid curve represents the fitting function described in the text.
Inset: a similar plot, for the critical {\it anisotropic} ABS model.
\vspace{1em}

%cp3crev.eps
\noindent FIG. 7.  Spread of activity in a typical realization of
the critical CP3 model ($q=0.63525$).
\vspace{1em}

%abscrev.eps
\noindent FIG. 8.  Spread of activity in a typical realization of
the critical ABS model ($q=0.66701$).
\vspace{1em}

%qc.eps
\noindent FIG. 9. 
Position $q_s$ of the singularity in the stationary probability density
of the extremal CP.  The upper line of singularities, $x=q$,
bifurcates from $q_s$ at the critical value $q_{c,E}$.
\vspace{1em}

%pcpe.eps
\noindent FIG. 10.
Stationary probability density $\overline{p}(x)$ in the CP$_E$ for
$q = 0.794 $ (left) and $q=0.85$ (right); system size $L=1600$.  
\vspace{1em}

%absancrev.eps
\noindent FIG. 11.
Spread of activity in a typical realization of
the critical anisotropic ABS model ($q=0.72370$).


\begin {thebibliography} {99}

\bibitem{BS}
	 P. Bak and K. Sneppen,   
	 Phys. Rev. Lett. {\bf 71}, 4083 (1993).

\bibitem{driven}
	 K. Sneppen,   
	 Phys. Rev. Lett. {\bf 69}, 3539 (1992).

\bibitem{leschhorn}
	 H. Leschhon and L.-H. Tang   
	 Phys. Rev. E {\bf 49}, 1238 (1994).

\bibitem{pmb94}
	 M. Paczuski, S. Maslov, and P. Bak, 
	 Europhys. Lett. {\bf 27}, 97 (1994).

\bibitem{pmb96}
	 M. Paczuski, S. Maslov, and P. Bak, 
	 Phys. Rev. E {\bf 53}, 414 (1996).

\bibitem{sornette95}
	D. Sornette, A. Johansen, and I. Dornic,
	J. Phys. I (France) {\bf 5}, 325 (1995).        

\bibitem{Grassberger:1995}
	 P. Grassberger, 
	 Phys. Lett. A {\bf 200}, 277 (1995).

\bibitem{sornette96}
	 D. Sornette and I. Dornic, 
	 Phys. Rev. E {\bf 54}, 3334 (1996).
     
\bibitem{gz96}
	 P. Grassberger and Y. C. Zhang,
	 Physica A {\bf 224}, 169 (1996).

\bibitem{BTW}
	 P. Bak, C. Tang and K. Wiesenfeld, 
	 Phys. Rev. Lett. {\bf 59}, 381 (1987).

\bibitem{Dickman et al:2000}
	 R. Dickman, M.A. Mu\~noz, A. Vespignani, and S. Zapperi,
	 Braz. J. Phys. {\bf 30}, 27 (2000).

\bibitem{flyvbjerg}
	 H. Flyvbjerg, K. Sneppen and P. Bak, 
	 Phys. Rev. Lett. {\bf 71}, 4087 (1993).

\bibitem{drossel02}
	    B. Drossel,
	    Adv. Phys. {\bf 50}, 209 (2001);
	    e-print: cond-mat/0101409.

\bibitem{harris}
	 T. E. Harris,
	 Ann. Probab. {\bf 2}, 969 (1974).

\bibitem{Garcia and Dickman:1}
	 G. J. M. Garcia and R. Dickman,
	 Physica A {\bf 332}, 318 (2004).

\bibitem{Garcia and Dickman:2003}
	 G. J. M. Garcia and R. Dickman, Physica A, to appear.
	 e-print: cond-mat/0408164.

\bibitem{Head and Rodgers:1998}
	 D. A. Head and G. J. Rodgers,
	 J. Phys. A {\bf 31}, 3977 (1998).

\bibitem{janssen}
	 H. K. Janssen, Z. Phys. B {\bf 42}, 151 (1981).

\bibitem{grassberger}
	    P. Grassberger, 
	    Z. Phys. B  {\bf 47}, 365 (1982).

\bibitem{liggett}
	 T. M. Liggett,
	 {\it Interacting Particle Systems},
	 (Springer-Verlag, New York, 1985).

\bibitem{hinrichsen} 
	 H. Hinrichsen,
	 Adv. Phys. {\bf 49}, 815 (2000).

\bibitem{konno}
	 N. Konno,
	 {\it Phase Transitions of Interacting Particle Systems},
	 (World Scientific, Singapore, 1994).

\bibitem{marro}
	J. Marro and R. Dickman,
	{\em Nonequilibrium Phase Transitions in Lattice Models}
	(Cambridge University Press, Cambridge, 1999).

\bibitem{jovanovic}
	 B. Jovanovi\'c, S. V. Buldyrev, S. Havlin and H. E. Stanley,
	 Phys. Rev. E {\bf 50}, R2403 (1994).

\bibitem{iwanpcp}
	I. Jensen,
	Phys. Rev. Lett. {\bf 70}, 1465 (1993).

\bibitem{dcp}
	    I. Jensen and R. Dickman, 
	    J. Phys. A{\bf  26}, L151 (1993). 

\bibitem{pmb95}
	 M. Paczuski, P. Bak, and S. Maslov, 
	 Phys. Rev. Lett {\bf 74}, 4253 (1995).

\bibitem{pmpreprint}
	 P. M. C. Oliveira, personal communication and preprint,
	 to appear in Braz. J. Phys.

\bibitem{pmnote}
	 The idea of using a single list, of sites with $x < 0.668$ (i.e.,
	 below a value slightly above the threshold), is already suggested
	 in Ref. \cite{pmpreprint}.

\bibitem{torre}
	 P. Grassberger and A. de la Torre,
	 Ann. Phys. (N.Y.) {\bf 122}, 373 (1979).

\bibitem{vendruscolo}
	    M. Vendruscolo, P. De Los Rios, and L. Bonesi
	    Phys. Rev. E {\bf 54}, 6053 (1996).

\bibitem{Maslov:1998}
	 S. Maslov, P. De Los Rios, M. Marsili, and Y.-C. Zhang,
	 Phys. Rev. E {\bf 58}, 7141 (1998).

\bibitem{head}
	 D. Head,
	 Eur. Phys. J. B {\bf 17}, 289 (2000).

\bibitem{Garcia and Dickman:2}
	 G. J. M. Garcia and R. Dickman, Physica A, to appear;
	 eprint: cond-mat/0403036.

\bibitem{jaff85}
        See
	J. Kamphorst Leal da Silva and M. Droz, 
	J. Phys. C {\bf 18}, 745 (1985), and references therein.

\bibitem{paritycons}
	P. Grassberger, F. Krause, and T. von der Twer, 
	J. Phys. A {\bf 17}, L105 (1984);
	P. Grassberger,
	{\it ibid}. {\bf 22}, L1103 (1989);
	H. Takayasu and A. Yu. Tretyakov, 
	Phys. Rev. Lett. {\bf 68}, 3060 (1992);
	I. Jensen, 
	Phys. Rev. E {\bf 50}, 3623 (1994).

\bibitem{mendes}
	 J. F. F. Mendes, R. Dickman, M. Henkel, and M. C. Marques,
	 J. Phys. A {\bf 27}, 3019 (1994).

\bibitem{cardy}
	 J. L. Cardy and R. L. Sugar,
	 J. Phys. A {\bf 13}, L423 (1980).

\bibitem{Boettcher:2000}
	 S. Boettcher and M. Paczuski,
	 Phys. Rev. Lett. {\bf 84}, 2267 (2000).


\bibitem{deBoer:1994}
	 J. de Boer, B. Derrida, H. Flyvbjerg, A.D. Jackson, T. Wettig,
	 Phys. Rev. Lett. {\bf 73}, 906 (1994).

\bibitem{deBoer:1995}
	 J. de Boer, A.D. Jackson, T. Wettig,
	 Phys. Rev. E {\bf 51}, 1059 (1995).

\bibitem{jensen96}
	 I. Jensen,
	 J. Phys. A {\bf 29}, 7013 (1996).

\end {thebibliography}
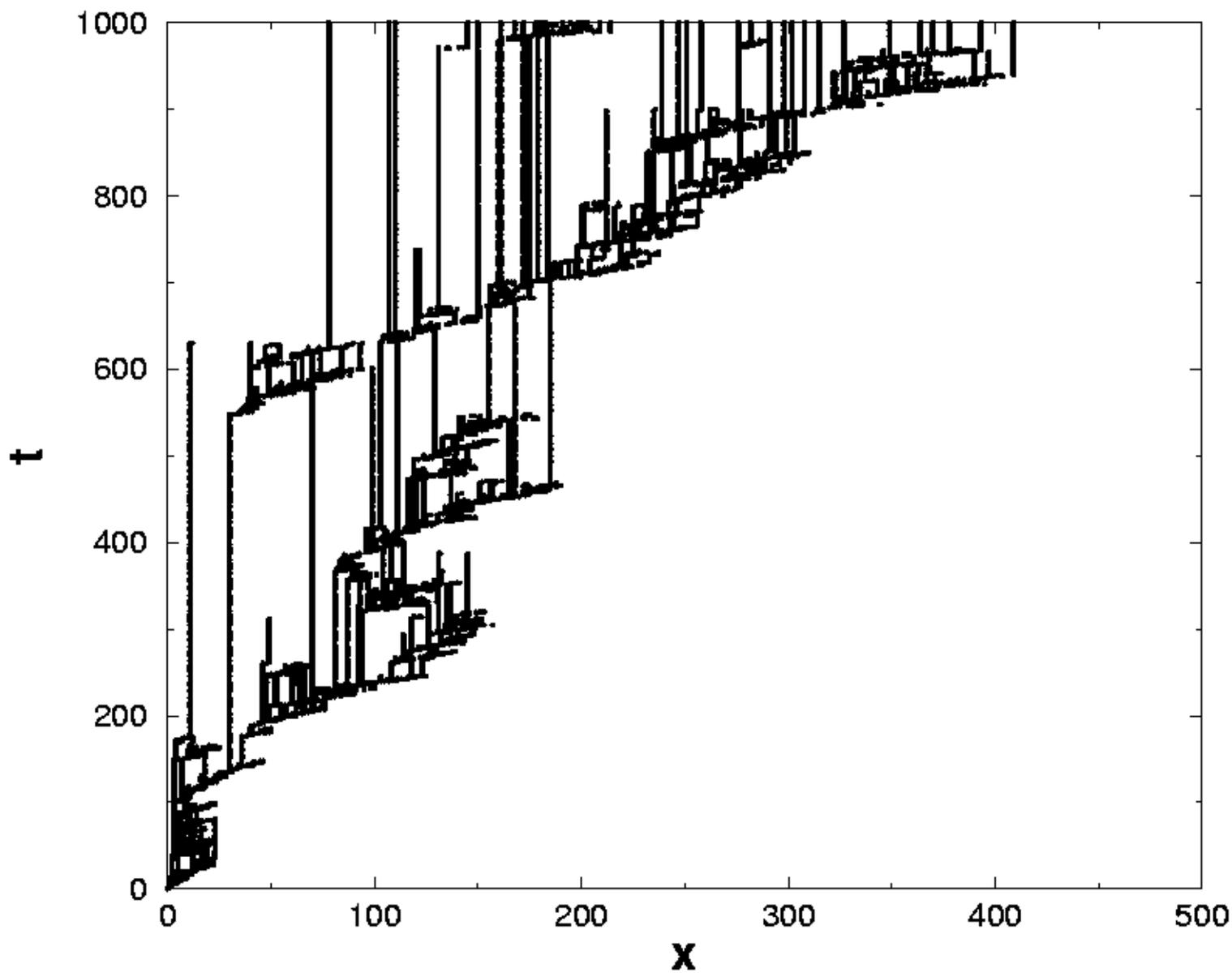
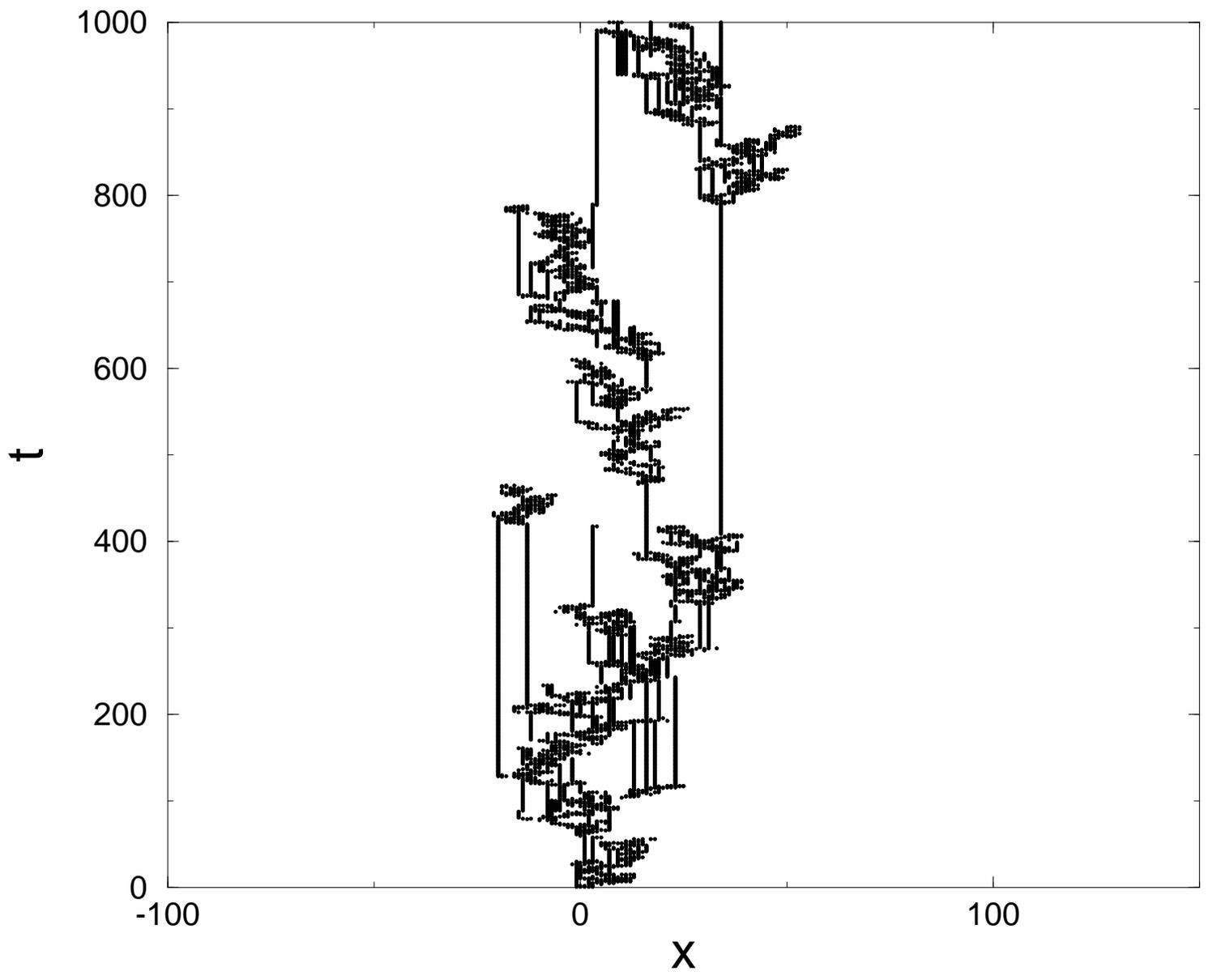
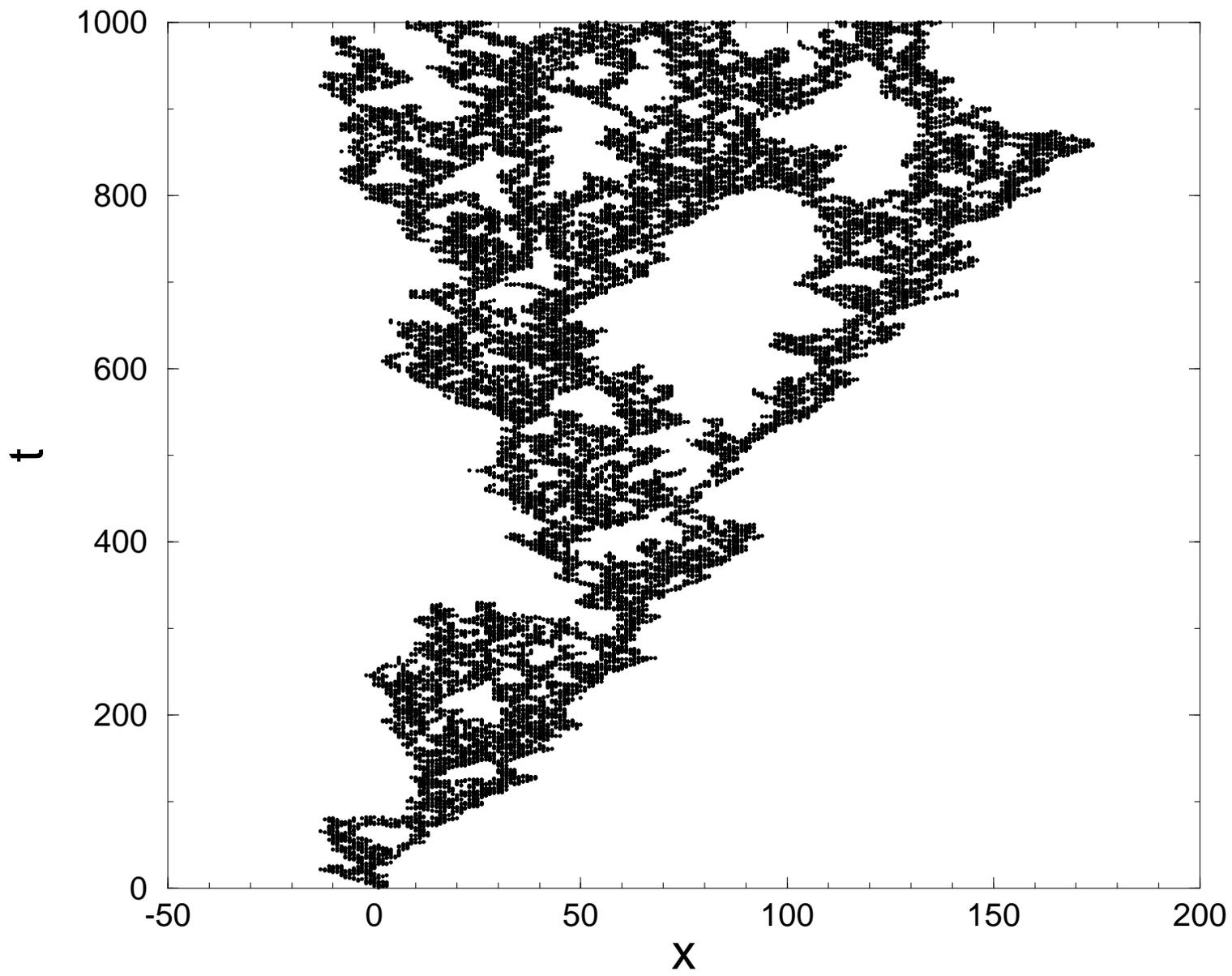
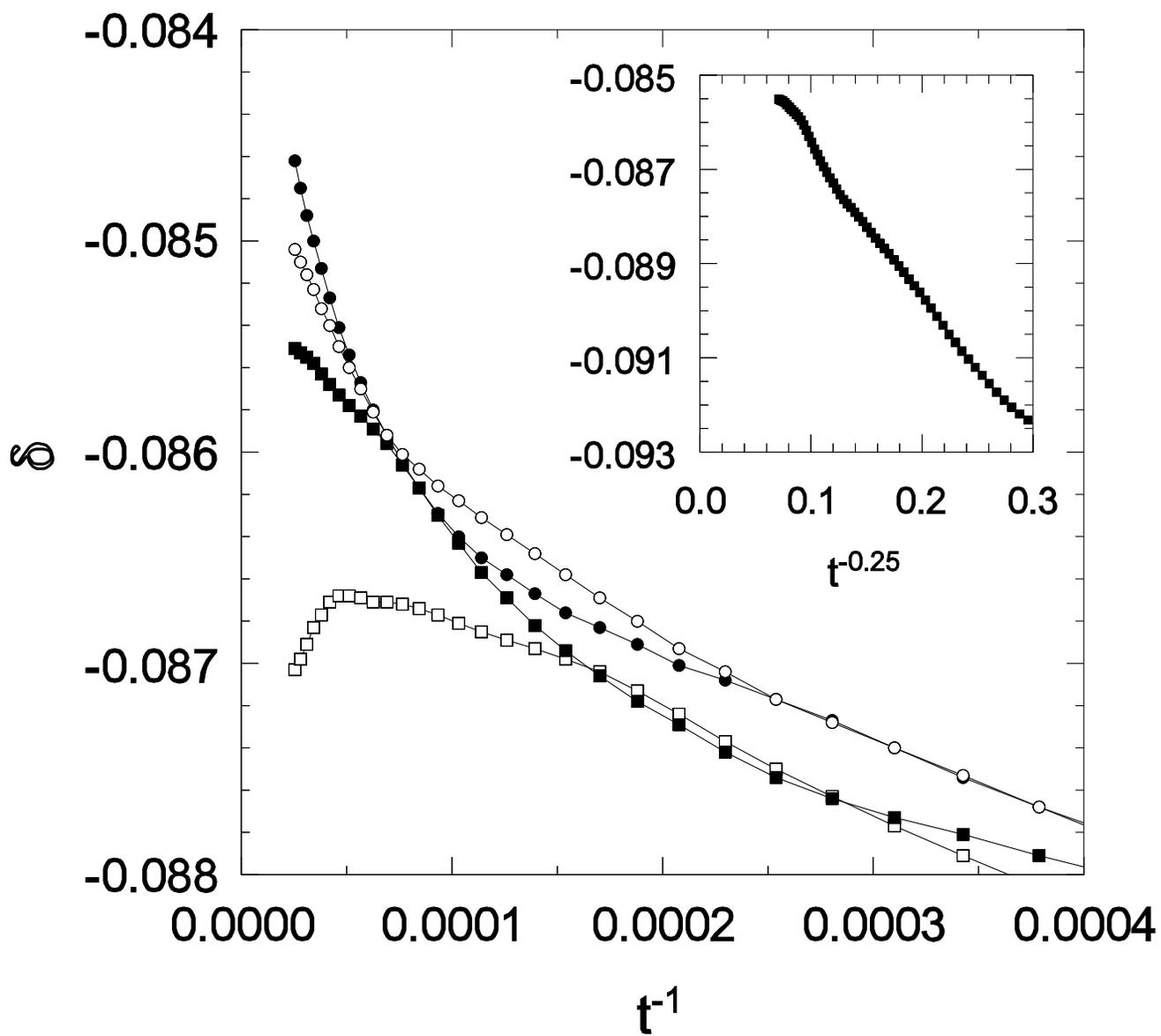
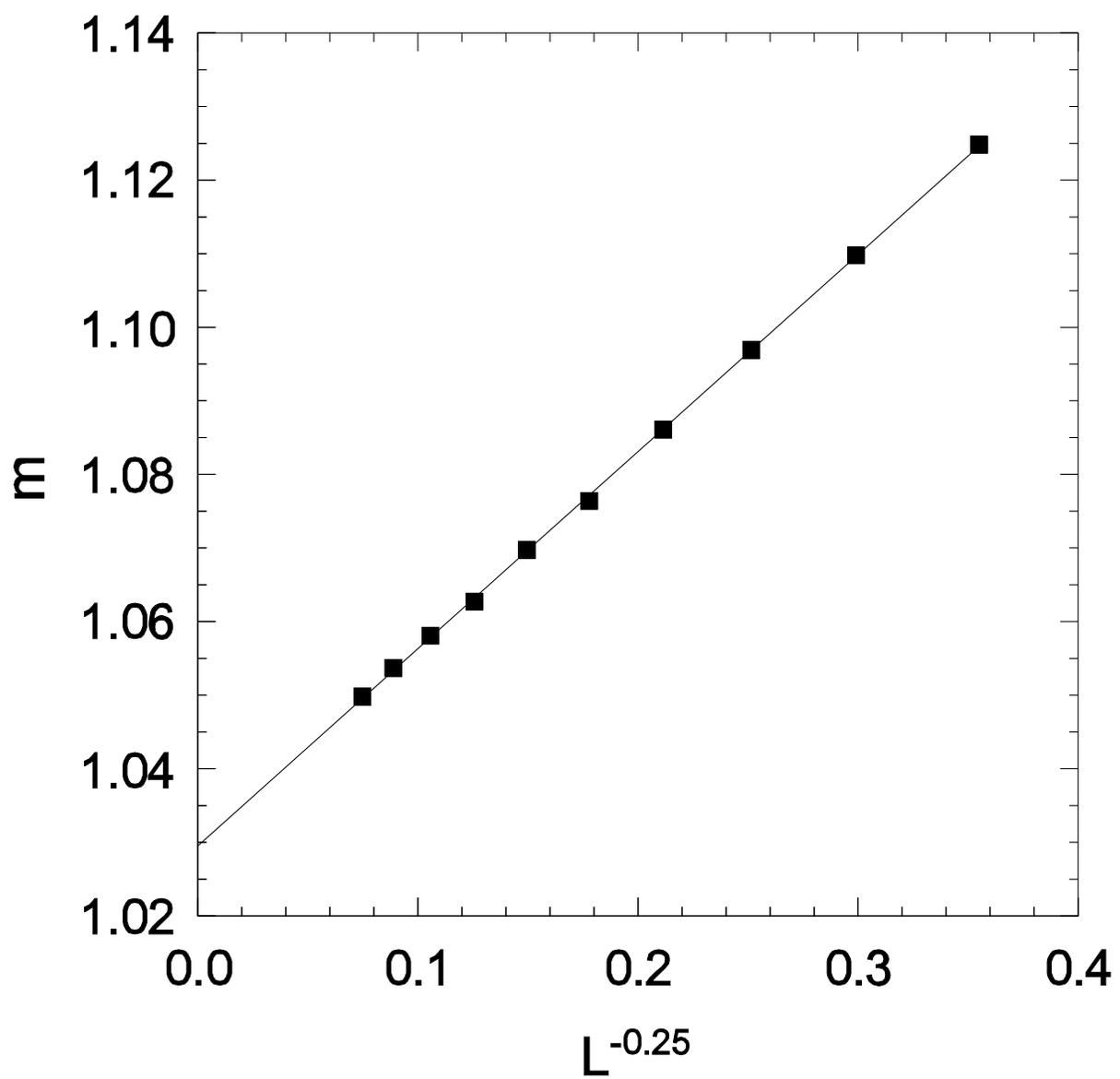
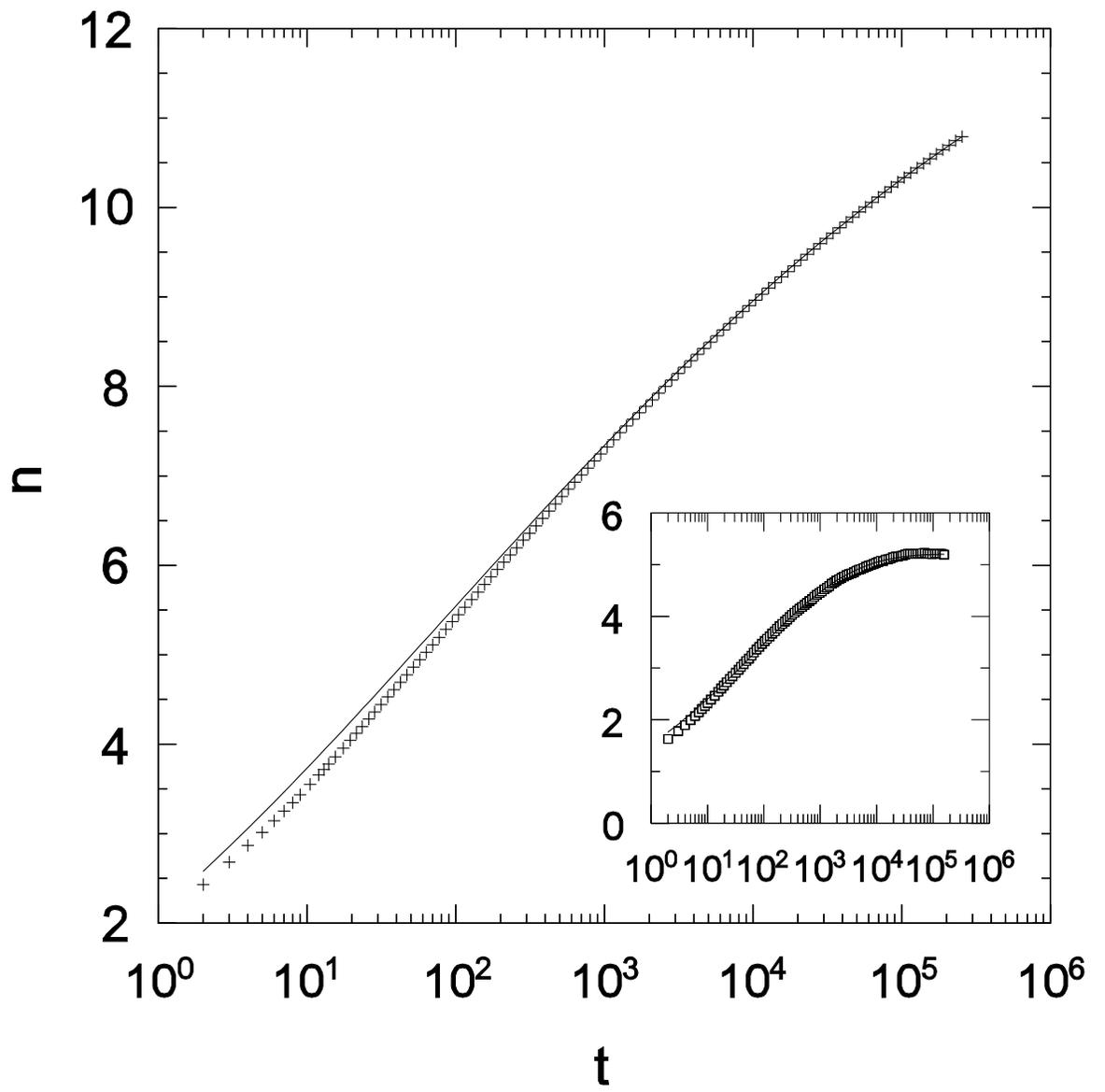
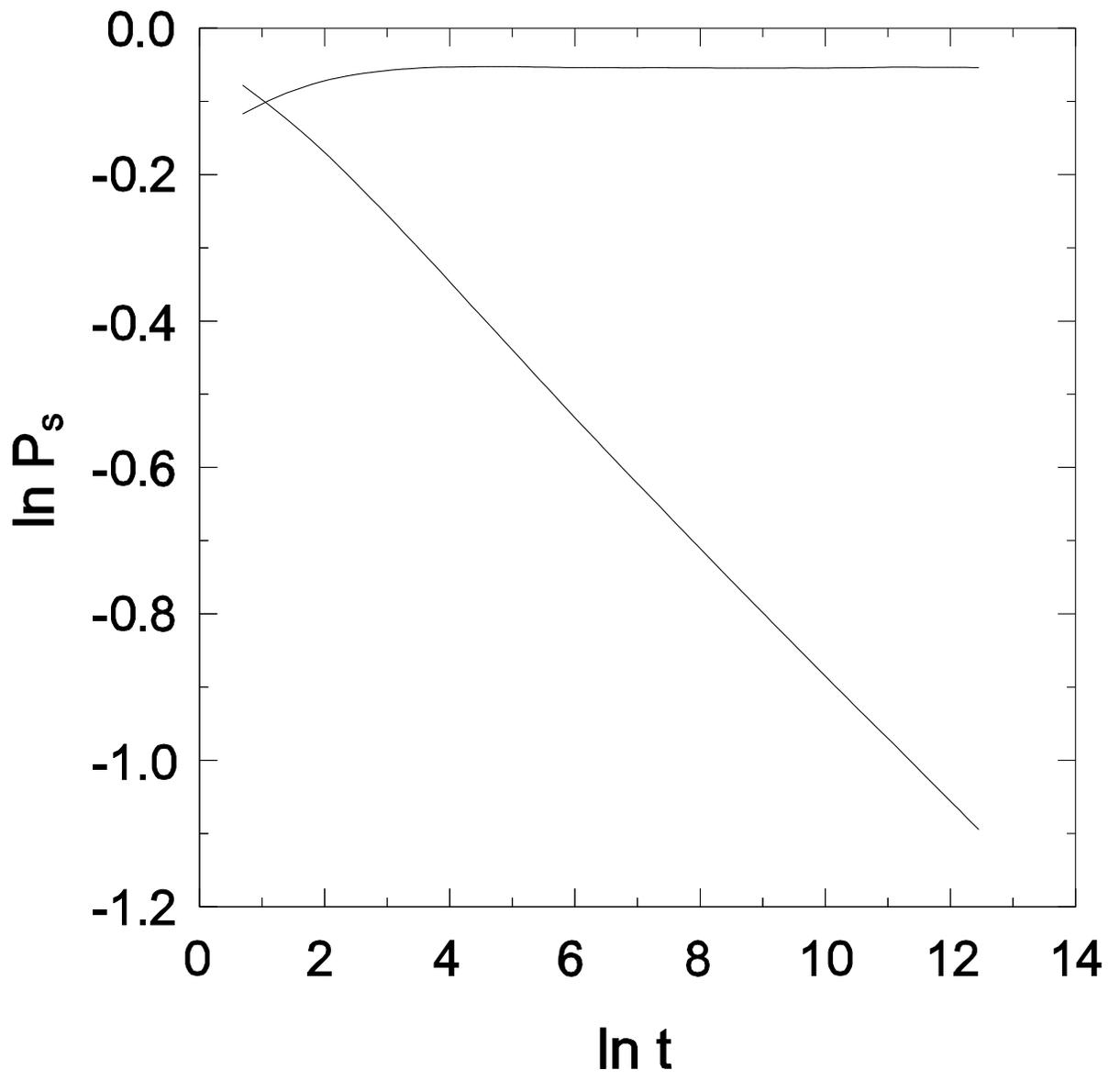
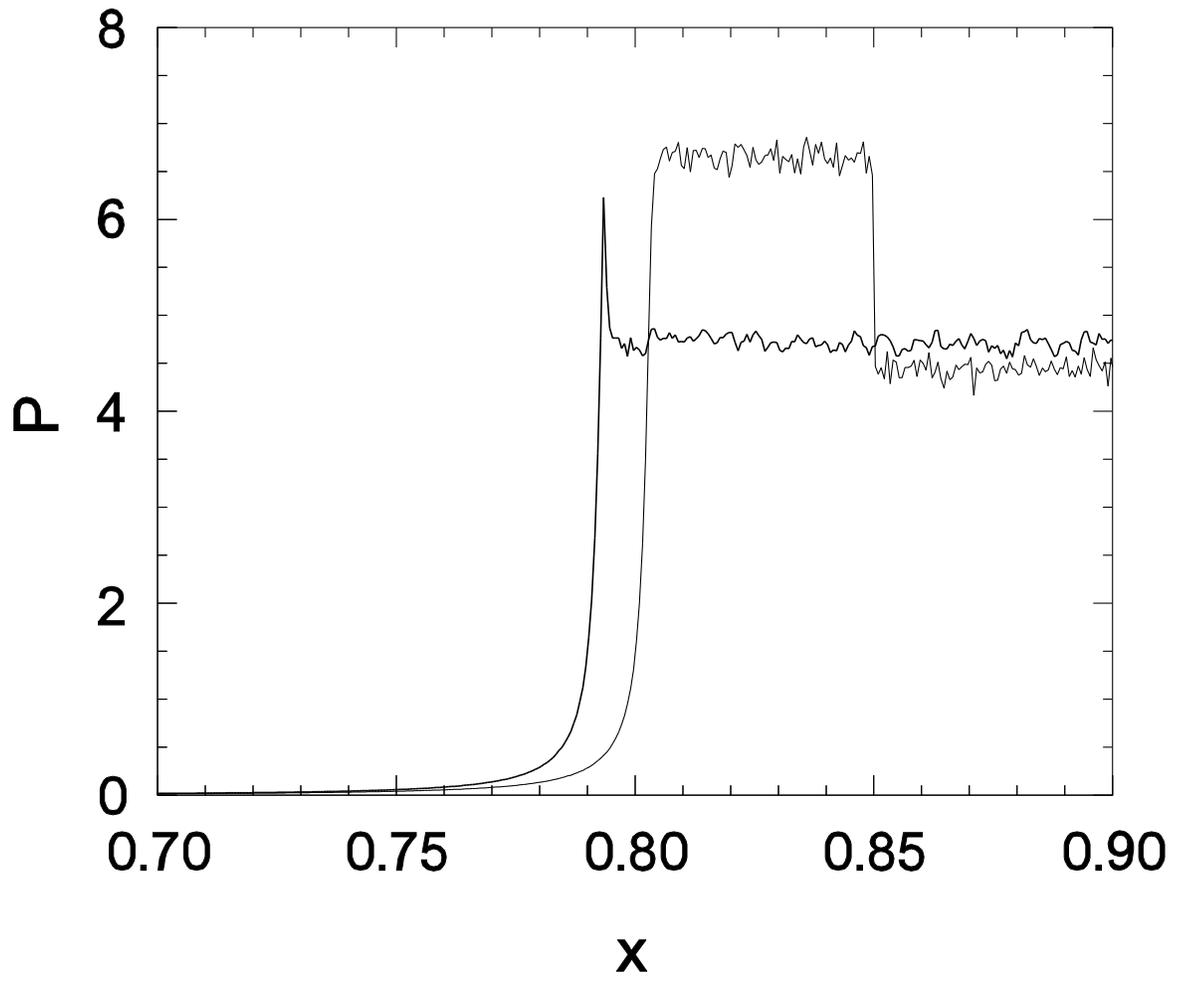
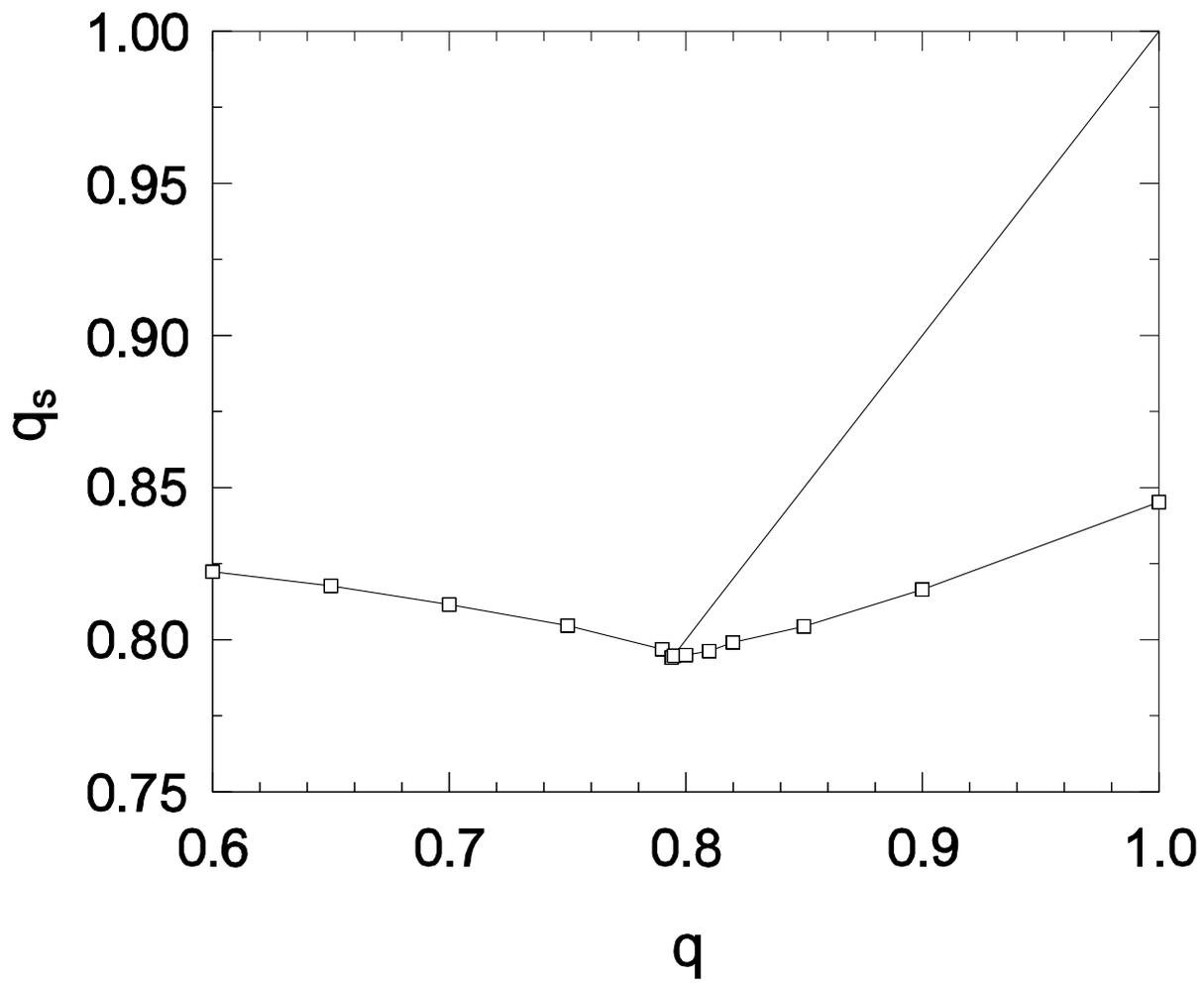
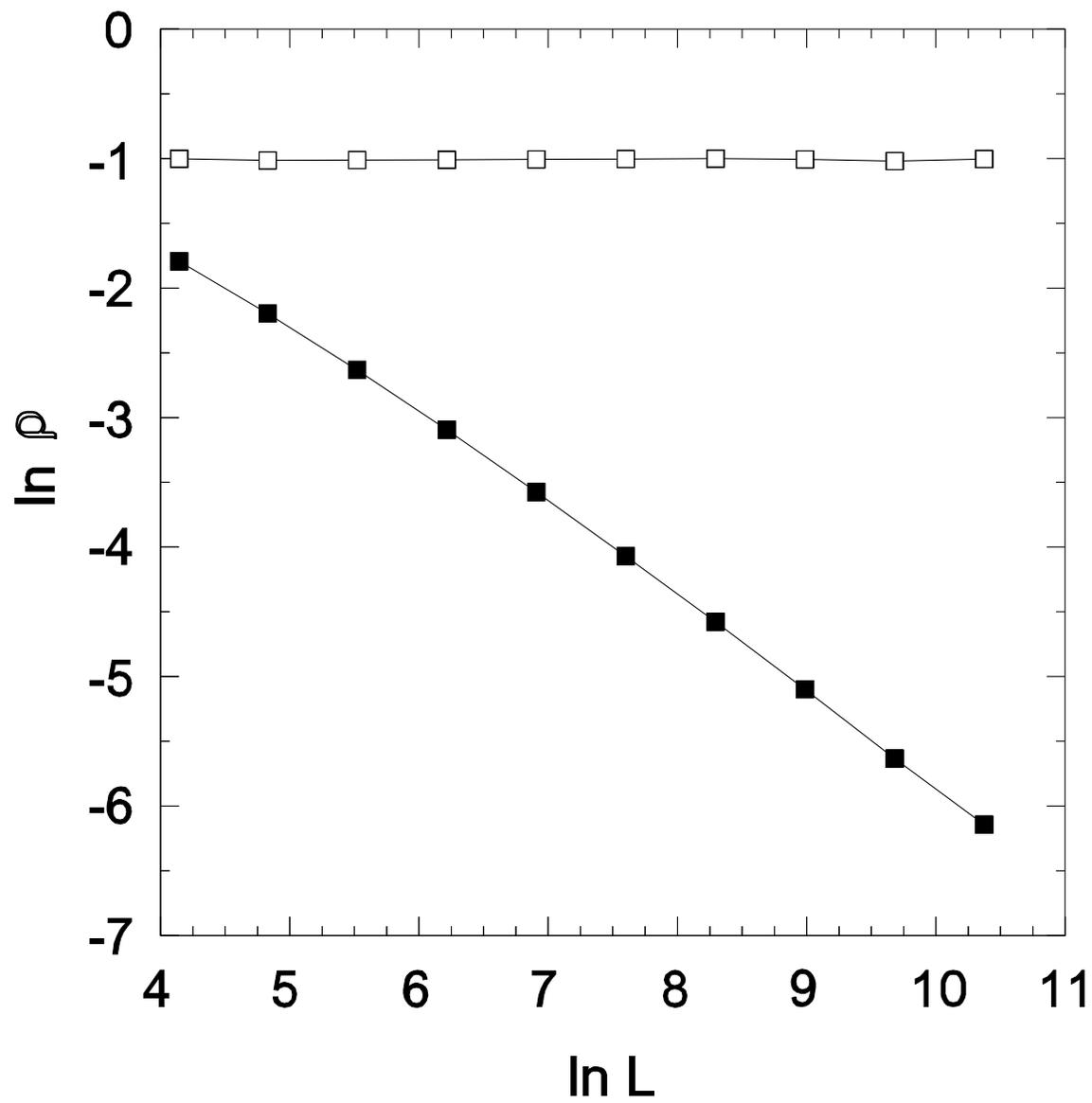
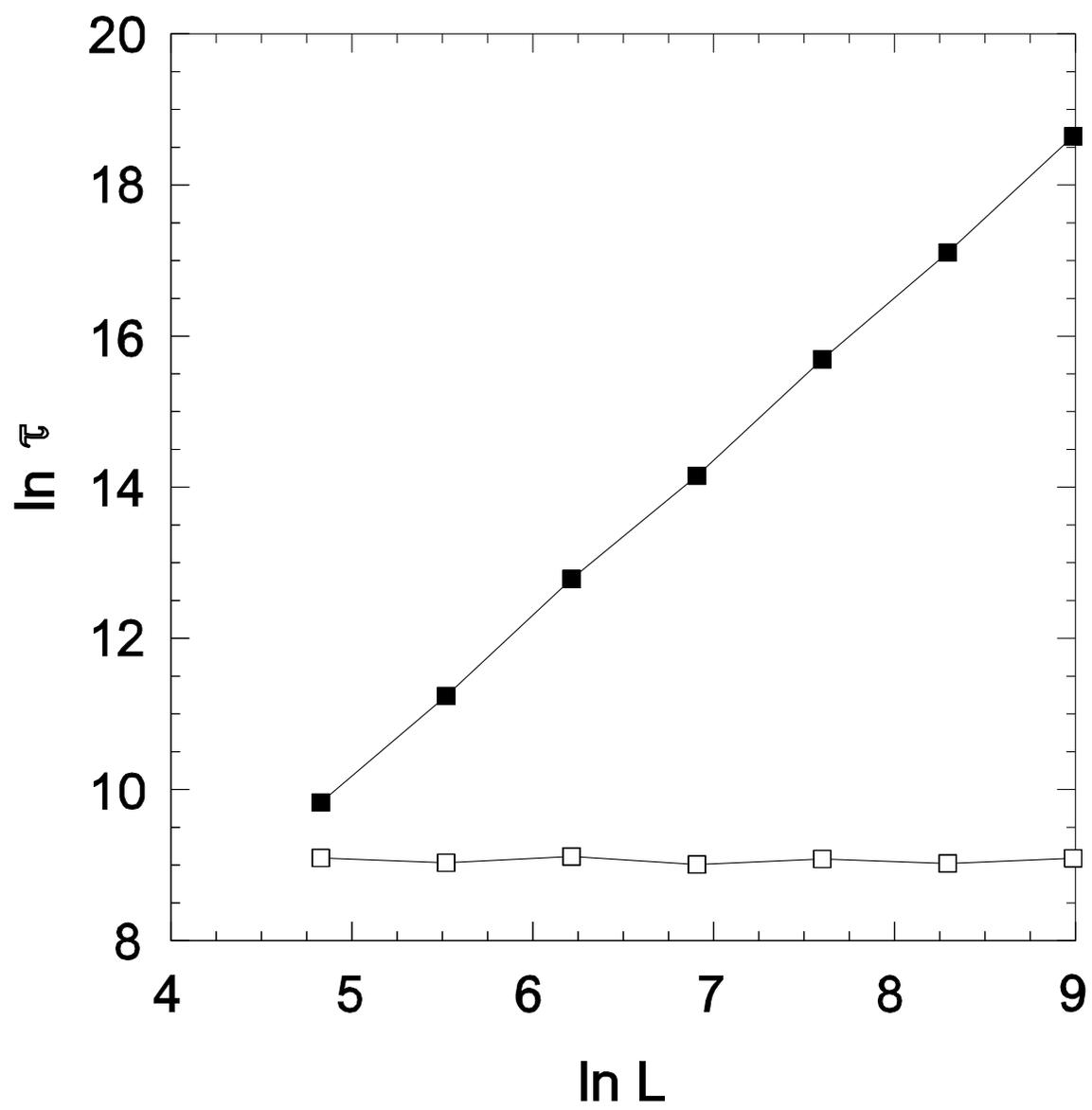
\end{document}